\documentclass[preprint2]{aastex} 

\newcommand{\noprint}[1]{}

\usepackage[nolists, tablesfirst, nomarkers]{endfloat}
\begin{document}

\title{Long-term variability of radio-bright BL Lacertae objects}

\author{Elina Nieppola, Talvikki Hovatta, Merja Tornikoski}
\affil{Mets\"ahovi Radio Observatory, TKK, Helsinki University of Technology}
\affil{Mets\"ahovintie 114, 02540 Kylm\"al\"a, Finland}
\email{elina.nieppola@tkk.fi}

\author{Esko Valtaoja}
\affil{Tuorla Observatory}
\affil{V\"ais\"al\"antie 20, 21500 Piikki\"o, Finland}
\affil{Dept. of Physical Sciences, University of Turku, 20100 Turku, Finland}
\and
\author{Margo F. Aller, Hugh D. Aller}
\affil{Department of Astronomy, University of Michigan}
\affil{Ann Arbor, MI 48109, USA} 


\begin{abstract}Radio-bright BL Lacertae objects (BLOs) are typically very variable and exhibit prominent flaring. We use a sample of 24 BLOs, regularly monitored at Mets\"ahovi Radio Observatory, to get a clear idea of their flaring behavior in the radio domain and to find possible commonalities in their variability patterns. Our goal was to compare the results given by computational time scales and the observed variability parameters determined directly from the flux curves. Also, we wanted to find out if the BLO flares adhere to the generalized shock model, which gives a schematic explanation to the physical process giving rise to the variability. We use long-term monitoring data from 4.8, 8, 14.5, 22, 37, 90 and 230 GHz, obtained mainly from University of Michigan and Mets\"ahovi Radio Observatories. The structure function, discrete correlation function and Lomb-Scargle periodogram time scales, calculated in a previous study, are analyzed in more detail. Also, we determine flare durations, rise and decay times, absolute and relative peak fluxes from the monitoring data. We find that radio-bright BLOs demonstrate a wide range of variability behavior, and few common denominators can be found. BLOs include sources with fast and strong variability, such as OJ 287, PKS 1749+096 and BL Lac, but also sources with more rolling fluctuations like PKS 0735+178. The most extreme flares can last for up to 13 years or have peak fluxes of approximately 12 Jy in the observer's frame. When the Doppler boosting effect is taken into account, the peak flux of a flare does not depend on the duration of the flare. A rough analysis of the time lags and peak flux evolution indicates that, typically, BLO flares in the mm -- cm wavelengths are high-peaking, i.e., are in the adiabatic stage. Thus, the results concur with the generalized shock model, which assigns shocks travelling in the jet as the main cause for AGN variability. Comparing the computational time scales and the parameters obtained from the flux curve analysis (i.e., rise and decay times and intervals of the flares) reveals that they do have a significant correlation, albeit with large scatter. 
\end{abstract}


\shortauthors{Nieppola et al.}  \shorttitle{Radio variability of BLOs}

\maketitle


\section{Introduction}

BL Lacertae objects (BLOs) are a relatively rare subclass of active
galactic nuclei (AGN). Traditionally the defining properties of BLOs
include a featureless optical spectrum, a flat radio spectrum and
vigorous variability at all frequency bands \citep[][and references
therein]{stein76, kollgaard94, jannuzi94, urry95}. Most BLOs are
thought to be highly beamed objects \citep{blandford79}, which means
that the relativistic jets emanating from the core are pointing very
closely at our direction. This is partly the cause of the featureless
optical spectrum; the non-thermal continuum emission from the jet can
swamp the thermal emission, including the emission lines, from the
host galaxy. However, in the case of less beamed objects, the lineless
spectrum must be created by other mechanisms.

The first BLO samples were discovered in either radio
\citep{stickel91, stickel93} or X-ray surveys \citep{gioia90,
stocke91}. Lately, surveys at different wavelengths \citep{londish02}
and the cross-correlation of existing radio and other waveband 
catalogs 
\citep{perlman98,caccianiga99, landt01, giommi05, turriziani07, plotkin08} have
produced a large number of new members to the class. Typically the
selection criteria are different from those of the pioneering surveys,
which has widened the definition of a BL Lac object; hence, the
variability characteristics of many of the newest BLOs are unknown due
to lack of data.

In a recent paper \citep{nieppola07}, 37 GHz fractional variability
indices from Mets\"ahovi Radio Observatory data were calculated for 90
BLOs. All sources, for which even a crude estimation of variability
could be calculated, exhibited an increase of 10 \% of the minimum
flux level at some point during the 3.5 year observation
period. Almost half of them doubled their minimum flux
density. However, this sample of 90 sources was only one fourth of the
full Mets\"ahovi BLO sample; the rest are too faint to allow any
variability analysis.

The variability of BLOs, like all flaring AGN, is thought to be caused
by shocks forming and travelling in the jet. The origin and early
development of these shocks is not yet very well understood. Once
propagating downstream in the jet, their evolution is easier to model
with Compton, synchrotron and adiabatic losses \citep{mar:gear85,
hughes89}. \citet{valtaoja92_moniIII} constructed a generalized view
of the Marscher \& Gear shock model, containing a general scenario of
the AGN flare behavior, to be used in comparison with
observations. The generalized model describes how the shape of the
shock spectrum remains unchanged as its peak moves from higher to
lower frequencies. The evolutionary track of the shock consists of
growth ($S_m\propto\nu^a_m$), plateau ($S_m\approx\textrm{constant}$),
and decay ($S_m\propto\nu^b_m$) stages, where $S_m$ and $\nu_m$ are
the turnover flux and frequency for the shock spectrum, and $a$ and
$b$ are model-dependent parameters. The flares can ideally be divided
into two groups: i) low-peaking flares, which will reach their maximum
intensity at lower frequencies than the observing frequency, and ii)
high-peaking flares, which have peaked at high frequencies relative to
the observing frequency and are already decaying. The observing
frequency, however, is not a constant quantity, but can be chosen
freely. Thus the low- and high-peaking classes are not fixed either:
the same flare can be low-peaking in one frequency band and
high-peaking on another. Consequently, the classes are not separated,
but rather opposite ends of a continuum of cases. Low- and
high-peaking flares should be distinguishable from observations. For
high-peaking flares the peak fluxes and time lags are strongly
frequency-dependent, the highest observing frequency peaking first
with the highest peak flux. For low-peaking flares the peaks are
nearly simultaneous in all frequencies, and the peak flux of the flare
is not significantly dependent on the frequency.

In this work we will study the long-term radio variability of BL
Lacertae objects at several frequencies. We will focus on a sample of
24 BLOs. The current number of definite and probable BLOs is over 1000
\citep{veron06}. This means that our sample is not a representative
cross-section for the whole population, but rather represents only the
rare, radio luminous BLOs. The bases of our work are the extensive
databases of Mets\"ahovi and University of Michigan Radio
Observatories at frequencies 4.8, 8, 14.5 GHz (UMRAO) and 22 and 37
GHz (Mets\"ahovi).

We will study the variability of our BLO sample from two points of
view: computational time scales obtained using statistical analysis methods 
and observed parameters of the flares,
e.g. the duration, rise and decay times, and absolute and relative
peak fluxes. The goal is to determine how well the computational time
scales correspond to the behavior we observe in the source, as well as
to gain a deeper understanding of what kind of variability can be
expected from radio-luminous BLOs, when monitored for tens of
years. We are also interested in how well the BLO flares adhere to the
generalized shock model of \citet{valtaoja92_moniIII}, and whether the
flares are mostly high- or low-peaking. A similar study is performed
on a larger AGN sample, including radio-loud quasars, in an
accompanying paper \citep{hovatta08b}.

In \S\ref{sam} we will present our sample and data. We briefly
describe the methods we used in \S\ref{met}. In \S\S\ref{ts} and
\ref{flares} we discuss the time scales and the observed flux curves,
respectively. In \S\ref{corr} we examine the correspondence between
the time scales and observed flaring, and describe the behavior of the
sample sources individually. We will finish with a discussion and
conclusions in \S\S\ref{dis} and \ref{con}, respectively. Throughout
this paper, we assume
$H_{0}=72\,\textrm{km}\,\textrm{s}^{-1}\,\textrm{Mpc}^{-1}$, 
$\Omega_m=0.27$, and $\Omega_{\Lambda}=0.73$.

\section{Sample and data}
\label{sam}

Our sample has been selected from the BLO sample monitored at the
Mets\"ahovi Radio Observatory for more than 20 years. The whole
Mets\"ahovi BLO sample comprises 398 sources, selected mainly from
\citet{veron00}. Most of the sources in the full sample are usually
very faint or non-detectable (meaning $S/N\leq4$) in the high radio
frequencies \citep{nieppola07}. The sample in this study contains
the very brightest sources with well-sampled flux curves. The
selection criterion was ample data from a period of at least ten years
in at least two radio frequencies. The sampling has to be sufficient
to determine the peaks of possible flares with adequate
accuracy. There are 24 available sources, 13 of which had one or
several significant flares during the observing period. 2 of the BLOs
are high-energy BLOs (HBLs), 4 intermediate BLOs (IBLs) and 18
low-energy BLOs (LBLs) \citep[for the basis of this classification,
see][]{nieppola06}. The source sample is listed in
Table~\ref{sample}. Columns (1) and (2) give alternative names for the
source, Cols. (3) and (4) give the right ascension and declination,
respectively, Col (5) gives the redshift, Col (6) gives the 
Doppler boosting factor of the source taken from \citet{hovatta08c}, 
and Col (7) gives the BLO
class according to \citet{nieppola06}. Column (8) indicates whether
the source has been included in the flare analysis in this work, and 
Col (9) gives the reference for the redshift.

\placetable{sample}

We used seven different frequency bands in the analysis, covering the
radio domain quite extensively. Low frequency data at 4.8, 8 and 14.5
GHz are from University of Michigan Radio Observatory. Details of the
observing system and data reduction can be found in
\citet{aller85}. 22, 37 and 87 GHz data are from Mets\"ahovi Radio
Observatory \citep{salonen87, terasranta92, terasranta98,
terasranta04, terasranta05, nieppola07}. The data reduction is
described in \citet{terasranta98}. The high frequency data at 90 and
230 GHz were obtained at the Swedish - ESO Submillimetre Telescope
(SEST) in La Silla, Chile, from 1987 to 2003 \citep[][and some
unpublished data]{tornikoski96_sest}, and also collected from the
literature \citep{steppe88, steppe92, steppe93, reuter97}. The 87 GHz
archival data from Mets\"ahovi were combined with the 90 GHz data to
form the 90 GHz flux curve.

\section{Methods}\label{met}
\subsection{Time scales}

The long term time scales of our sample have been calculated in
\citet{hovatta07}, where our sample represented the BLO class in
comparison with other AGN subgroups. In this paper we report the
individual time scales of the BLOs.  The timescales have been
calculated in three ways: using the structure function (SF), the
discrete correlation function (DCF) and the Lomb-Scargle
periodogram. DCF and periodogram may provide several time scales of
different duration. In this paper we will concentrate only on the
most significant ones. The theoretical aspects of the methods are
discussed in \citet{hovatta07} and the references therein. In the
interest of comparing them with the observed flux curves, we have kept
the computational time scales in observer's frame throughout the paper
and have not performed any redshift or Doppler corrections on them.

The three methods ($T_{SF}$, $T_{DCF}$, and $T_{P}$) respond to flux
variations of different scales. SF is the most sensitive, picking up
the short time scales and the structure of the flares, like the rise
and decay times. DCF and L-S periodogram are meant to provide the time
scale of longer variations, like the peak--to--peak intervals of major
flares. The periodogram was originally developed to search for strict,
sinusoidal periodicities. In this context it is used to give a
characteristic time scale, and a result from the periodogram analysis
does not mean that the source is strictly periodic. The distinctions of
SF, DCF and periodogram are also thoroughly discussed in
\citet{hovatta07}.

\subsection{Flare parameters}\label{met_flare}

For comparison with the computational time scales, we determined some
flare parameters directly from the observations. We use the word
"flare" to describe a separate period of heightened activity. We have
not separated the individual shocks contributing to the flux density
rise, so in some cases one flare may include several components but
this is more pronounced in the low-frequency domain. The start and end
times, and, thus, also the duration, of the flares are based on a
careful visual estimation of each flux curve. Their error is dependent
on the sampling frequency: with well-sampled flux curves the
determination of the flares is more accurate. In 90 and 230 GHz the
sampling was often too poor to allow the definition of the flares,
which is why these frequencies are, in many cases, at least partly
excluded from the analysis. A flare was included in the analysis if it
was discernible at least at two frequencies, one of them being 22 or
37 GHz. The absolute peak flux and peak time are defined
straightforwardly by the highest flux density measurement between the
start and end times of the flare. The relative flux is defined as the
difference between the flux minimum at the start of the
flare and the absolute peak flux of the flare. The rise time is the
interval between flare start and the time of the peak flux, while the
decay time is the interval between the peak time and the end of the
flare, defined as the flux minimum after the flare. 
No flare had a significant plateau stage. Unless clearly stated
otherwise (see \S~\ref{mor}), the flare parameters are also discussed
in the observer's frame.

\section{Time scales}\label{ts}

 The time scales are available for all frequencies that had a
 sufficiently well-sampled flux curve for their definition and are
 listed in Table~\ref{timescales}. In many cases, only one or two
 methods could be utilized. There were two factors which complicated
 the determination of the time scales. Firstly, some sources are very
 faint and exhibit a relatively uneventful flux curve. In that case,
 the errors of the flux measurements are large compared to the flux
 densities. This leaves a considerable margin of error in the modest
 flux rises and falls, and adds to the uncertainties of the time
 scales. Secondly, some objects exhibit rapid variability and
 outbursts of different magnitudes, which results in many different
 time scales obtained in DCF and periodogram analyses. In the DCF
 analyses, we have chosen the first discernible peak, after the DCF
 has been on the negative side, as the representative time scale. In
 the case of the periodogram, the highest peak represents the most
 significant time scale. If such a peak occurred at a time scale that
 was longer than half of the total observing period, it was discarded
 to avoid spurious time scales. In some cases, the most significant
 DCF time scale is, for example, the second most significant according
 to the periodogram, in which case their values in
 Table~\ref{timescales} can be different.

\placetable{timescales}

\begin{figure}
\epsscale{0.5}
\plotone{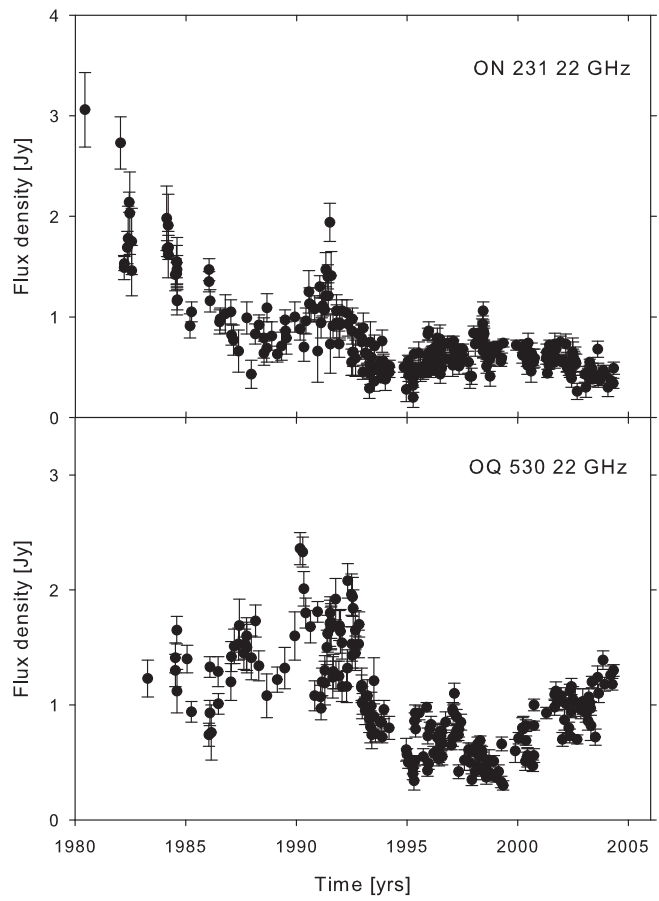}
\caption{Flux curves of ON 231 (top panel) and OQ 530 (bottom panel) at 22\,GHz}
\label{on231}
\end{figure}

There are some sources which have only a lower limit of $T_{SF}$
listed in Table~\ref{timescales}. In many cases, this lower limit time
scale is long. We do not consider these as definite; rather, they are
an indication that the determination of the time scale was difficult,
often due to uneventful flux curves. ON 231 (see for 
Fig. \ref{on231} top panel for the flux curve at 22 GHz) is a 
good example of how one
prominent flare can affect the determination of the time scales. It
underwent a relatively strong outburst in the late 1970's - early
1980's, and the decay stage has been recorded in all frequency bands
from 4.8 to 37 GHz. The more subtle flux variations are superposed on
a steady decline of flux, which lasted from the very beginning of
monitoring in the early 1980's to mid-1990's. While representing a
valid variability time scale itself, this slow fluctuation hinders the
definition of the shorter time scales. This affects mainly the SF,
which is designed to distinguish the shortest time scales of
variability, lengthening the time scale of these minor variations. As
a result, most of the SF time scales calculated for ON 231 give only
lower limits. Another example of a lower limit is the 22 GHz SF time
scale of OQ 530 (see for Fig. \ref{on231} bottom panel for the 
flux curve at 22 GHz). Although the flux curves at 22 and 37 GHz are very
similar, and time windows are comparable, the SF gives differing
results. However, the structure of the SF is very complicated. At 22
GHz, hints of shorter time scales can be seen, but their plateaus in
the SF are not clear enough to be picked as the representative time
scale. Also, the 90 GHz DCF time scale of 1308+326 has exceptionally
large errors, and the value should be treated with caution.

\placefigure{on231}

Studying the time scales of Table~\ref{timescales} more closely, we
find that three objects, namely S5 0716+714, Mark 421 and PKS
1749+096, are among the five sources with the shortest time scales,
independent of the method of calculation. Therefore, they can well be
dubbed the BLOs with the fastest variability. Their shortest time
scales vary from 70 days ($T_{SF}$ for S5 0716+714 at 14.5 GHz) to 2.8
years ($T_{P}$ for PKS 0754+100 at 37 GHz). It is noteworthy that Mark
421 has very low flux levels and the definition of the time scales
suffers, as explained earlier. Other short time scale sources include
OJ 287 (according to both SF and periodogram), PKS 0754+100, S4
0954+65, Mark 501, and 3C 371.0.

The objects 3C 446 and 1308+326 (according to both DCF and
periodogram), OJ 425 (according to both SF and periodogram), as well
as PKS 0735+178, ON 231, and 4C 14.60 seem to be good examples of
sources with long time scales and slow variability. They have longest
time scales ranging between 6.8 ($T_{SF}$ for ON 231 at 4.8 GHz) and
15.2 years ($T_{P}$ for 3C 446 at 8 GHz).

In general, the DCF and periodogram time scales have quite a good
correspondence, and the SF time scales are clearly shorter (for more
information on the correspondence of the time scales in general, see
\citet{hovatta07}). For some sources, however, the differences between
the time scales obtained with the three methods can be large. Usually
this is because the most significant time scale is defined differently for 
these methods. In most cases when the periodogram and DCF
time scales differ significantly, similar time scale to $T_{P}$ has
been seen also in the DCF but is has not been the most significant one.  
In some cases the different frequency bands have strikingly dissimilar values. 
This
is typically due to the faintness of the source and the low amplitude 
variability,
which can make the determination of a time scale a difficult task and
overemphasize the influence of discrepant datapoints in the flux
curve.

The mean values of $T_{DCF}$, $T_{SF}$ and $T_{P}$ for various AGN
subgroups, including BLOs, are reported in \citet{hovatta07}.

\section{Observed radio outbursts}\label{flares}

\subsection{Flare morphology}\label{mor}

\begin{figure}
\epsscale{0.5}
\plotone{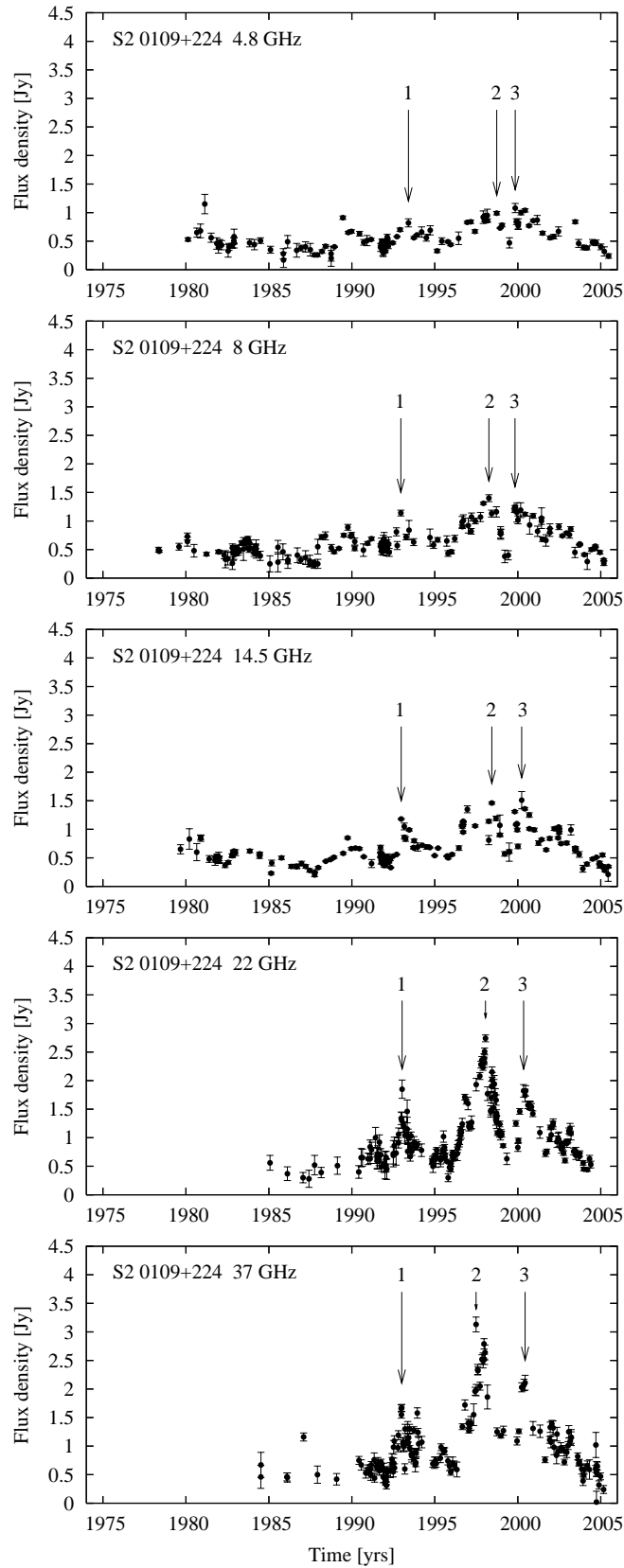}
\caption{Flux curves of objects in the flare analysis. The peak of each flare included in the analysis is marked in the curve. Plots for all sources are available in the electronic edition of the journal.}
\label{fluxcurves}
\end{figure}

There are 13 BLOs which exhibit significant flaring during our monitoring period. The flux curves of the flaring sources are available in Fig.~\ref{fluxcurves} (Figures \ref{fluxcurves}.2 -- \ref{fluxcurves}.13 are available in the online version of the journal), where each flare, identified at 22 or 37\,GHz, is marked. It is evident that the flux curves are very diverse in morphology.

\placefigure{fluxcurves}

\subsubsection{Sample means by source and frequency}\label{means}

The mean values of flare duration, rise time, decay time, absolute
peak flux and relative peak flux, determined as described in
\S\ref{met_flare} are listed in Table~\ref{source_means} for each of
the 13 sources. The parameters have been calculated as an average for
all frequency bands, and for 37 GHz separately for comparison. In one
end we have objects like AO 0235+164, OJ 287, PKS 1749+096 and BL Lac
itself with rapid and frequent fluctuations. In the other end of the
range we find PKS 0735+178, 1308+326 and 3C 446 which have flares that
last for several years, with only a couple of them covered by the span
of our observations. In fact, there is much doubt about the nature of
the latter objects. 1308+326 and 3C 446 were originally included in
our BLO master list because several authors have classified them as
borderline cases between BLOs and quasars \citep{gabuzda93, falomo94, 
laurent99, aller99}.
Later they have been listed as quasars in the Veron-Cetty \& Veron
Catalogs. While listed as BLO in \citet{veron06}, PKS 0735+178 exhibits a 
similar type of radio flux curve.

\placetable{source_means}

The typical BLO flare has a measured peak flux well below 10 Jy, as
seen in Table~\ref{source_means}. The average relative peak fluxes are
mostly below 5 Jy. The brightest flares in our sample are those of 3C
446, measured in both absolute and relative flux.

In Table~\ref{parameters} we present the minimum, maximum, mean and
median values of flare duration and absolute and relative peak fluxes
for each frequency used in our analysis. We calculated the 
relative peak fluxes in two different ways, first by subtraction 
($S_\mathrm{max}-S_\mathrm{min}$) and secondly by division 
($S_\mathrm{max}/S_\mathrm{min}$). The latter can be considered as a 
variability index for each flare. In the discussion of shock models 
(cf. Sect. \ref{dis}) the first one is used. The duration and absolute peak 
flux of the flares 
are also tabulated in two different ways. We show the absolute duration 
in years and absolute peak flux in janskys for all the frequency bands and 
also the values of each 
individual flare normalized to the value at 22\,GHz (in the calculation of 
the durations this was not possible 
for flare 2 of 1308+326 and flare 7 of BL Lac because their duration at 
22\,GHz could not be calculated).

There were 34 flares in total
at 4.8 GHz, 38 at 8 GHz, 45 at 14.5, 22 and 37 GHz, 17 at 90 GHz and 8
at 230 GHz. In duration, the difference between the minimum and
maximum values is vast. The majority of the flares are relatively
short in duration for radio band events, and sources PKS 0735+178 and
1308+326 alone have flares extending over 6 years, as we already
learned from Table~\ref{source_means}. This can be seen in the median
values in Table~\ref{parameters}, which are between 2.3 and 2.7
years. The duration of the flares changes little with frequency which 
can be clearly seen in the relative durations. The
absolute median values are slightly longer at 4.8 GHz and 
8 GHz than in the
higher frequencies, but only by 0.4 years at most.  However, at 90 and
230 GHz the sparser sampling of the flux curves does not allow as
accurate determination of the flare duration as the frequent sampling
of the lower frequencies. In reality, the mean duration of the 90 and
230 GHz flares may be slightly shorter.

The absolute peak flux exhibits a stronger correlation with
frequency. The flare peak fluxes range between 0.7 and 12.1 Jy. The
median values rise with frequency up to 5.1 Jy at 37 GHz which is 
also seen when the normalized absolute peak fluxes are studied. In 90 and
230 GHz, the median peaks are more moderate, 4.4 and 3.2 Jy,
respectively, corresponding to 90\% and 67\% of the flux at 22\,GHz.
. Also in this case the sparse sampling of the highest
frequencies has its effect: with more datapoints their median peak
fluxes might be higher. The parameters of the relative flux behave
roughly in the same manner. The relative fluxes of BLO flares range
between 0.4 and 10.1 Jy. It is also seen that maximum fluxes 
are 1.4 to 18.5 times higher than minimum fluxes.

\placetable{parameters}

\subsubsection{Flare intensity vs. duration}

\begin{figure}
\epsscale{0.5}
\plotone{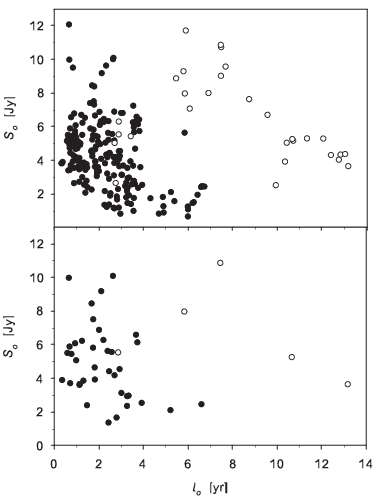}
\caption{The absolute peak flux plotted against the duration of the flare, with all flares included (top panel) and only 37 GHz flares included (bottom panel). Datapoints of typical BLOs are marked with black circles, those of the three quasar-like objects are marked with open circles.}
\label{flux_dur}
\end{figure}

In Fig.~\ref{flux_dur} we have plotted the absolute peak flux, $S_o$,
of each flare against the flare duration, $t_o$ using all frequencies
(top panel) and 37 GHz only (bottom panel). The distribution of the
datapoints seems to be bimodal, with a dividing line running from the
upper left corner to lower right. On a closer inspection we find that
the datapoints in the upper right are those of the quasar-like objects
PKS 0735+178, 1308+326 and 3C 446 (see \S\ref{means}). Their long and
intense outbursts thus clearly differ from the typical BLO
flares. Only one flare of 3C 446 is more BLO -like, as can be seen in
Fig.~\ref{flux_dur}.

\placefigure{flux_dur}

We also find a distinct declining trend in both panels of
Fig.~\ref{flux_dur}. It is evident in both the typical BLOs and
quasar-like objects, but less so in the latter, however, due to the
one weak and short flare of 3C 446. The correlation for the ``genuine''
BLOs is significant also according to the Spearman rank correlation
test. When all frequencies are considered, the Spearman correlation
coefficient is $\rho$= -0.238 and the probability of no correlation
$P$=0.001.  The strength of the correlation is slightly distorted by
the fact that all the frequencies are included in the calculation, and
thus every flare is counted for more than once. We also checked its
significance at 4.8, 8, 14.5, 22 and 37 GHz separately. In this case,
the significance of the correlation seems to vanish.

\begin{figure}
\plotone{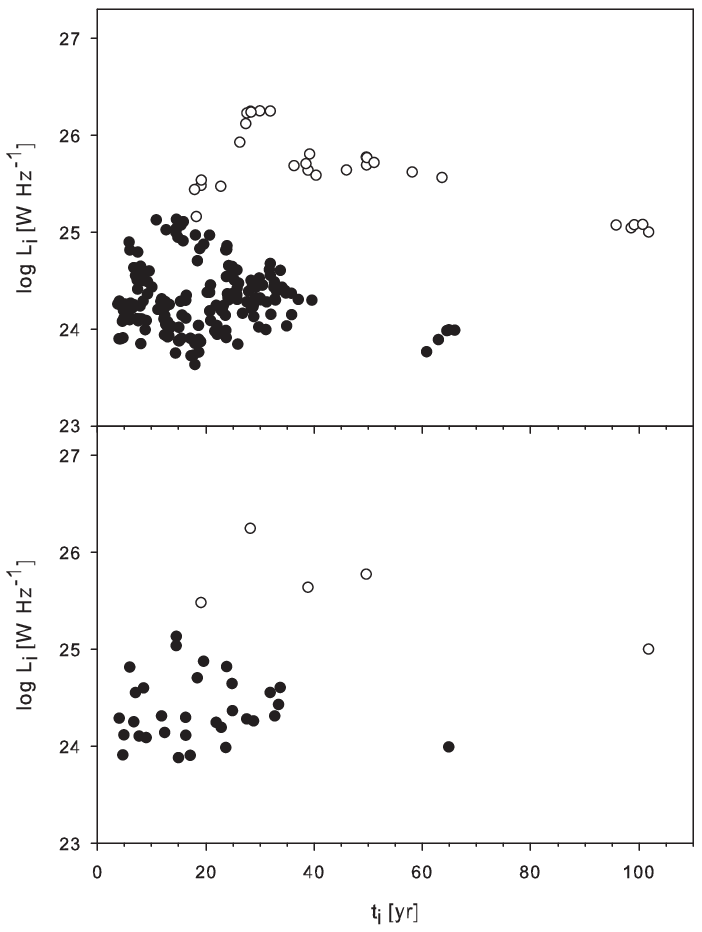}
\caption{The Doppler-corrected flare peak luminosity plotted against the Doppler-corrected duration of the flare, with all flares included (top panel) and only 37 GHz flares included (bottom panel). Datapoints of typical BLOs are marked with black circles, those of the three quasar-like objects are marked with open circles.}
\label{lumkorr}
\end{figure}

A natural explanation for the negative trend observed in
Fig.~\ref{flux_dur} is the effect of Doppler boosting. In 
Eq. \ref{doppler} the Doppler 
boosting factor is defined by the Lorentz factor of the jet flow $\Gamma$, 
speed of the jet $\beta$, and the viewing angle to the line of sight 
of the observer $\theta$.
\begin{equation}\label{doppler}
D = \frac{1}{\Gamma(1-\beta\cos\theta)}
\end{equation}
As the boosting
factor $D$ increases, the internal time scales of the source get
shorter and flux levels become higher. To better investigate the
intrinsic properties of the sources, we plotted the Doppler-corrected
peak luminosity, $L_i$, of each flare against the Doppler corrected
duration, $t_i$, of the flare (Fig.~\ref{lumkorr}). The corrections
and luminosity calculation were performed with equations (see, e.g.,
\citet{kembhavi99}, but note the typing error in their Eq.~(3.102))
\begin{equation}\label{dopdur}t_i=\left(\frac{D_{var}}{1+z}\right)
t_o\end{equation} and
\begin{equation}\label{doplum}L_i=\left(\frac{1+z}{D_{var}}\right)^{3+
\alpha} \frac{4 \pi d_L}{1+z} S_o \end{equation} where $z$ stands for
redshift and $d_L$ for the luminosity distance. The subscripts $i$ and
$o$ denote intrinsic and observational quantities, respectively. The
Doppler factors $D_{var}$ were taken from \citet{hovatta08c}, where they 
have been determined from our extensive database of total flux 
density observations at 22 and 37\,GHz in the same manner as in
\citet{lahteenmaki99_tfdIII}. Applying an exponential fit to individual 
shock components extracted from the flux curves gives the observed variability 
brightness temperature. Comparing it to the
intrinsic brightness temperature gives the amount of boosting. In
Eq.~\ref{doplum} we have assumed an evolving feature in the jet, in
keeping with the shock scenario, and $\alpha = 0$ ($F \propto
\nu^{-\alpha}$). For two sources in our flare analysis 
(S2 0109+22 and PKS 0422+004) no Doppler boosting factor was detemined in 
\citet{hovatta08c} due to uncertain redshifts, and these are not included 
in Figs.~\ref{lumkorr} and \ref{dec_rise}.

\placefigure{lumkorr}

Fig.~\ref{lumkorr} shows us that when the boosting effect is taken
into account, the data set looks very different. Only 1308+326 has a
strikingly long flare duration, followed closely by PKS 1413+135 at
lower luminosities. There is no correlation between $L_i$ and
$t_i$ among the ``genuine'' BLOs (black circles in Fig.~\ref{lumkorr}) 
when all frequency bands are included ($\rho$=0.0939 and
$P$=0.1095) or with 37 GHz datapoints only ($\rho$=0.1524 and $P$=0.1986). 
The difference between
Figs.~\ref{flux_dur} and ~\ref{lumkorr} attests again to the
substantial influence the boosting effects have on our observations of
AGN. The correlations are very similar using the relative flare
luminosity.

\subsubsection{Flare shapes}

\begin{figure}
\plotone{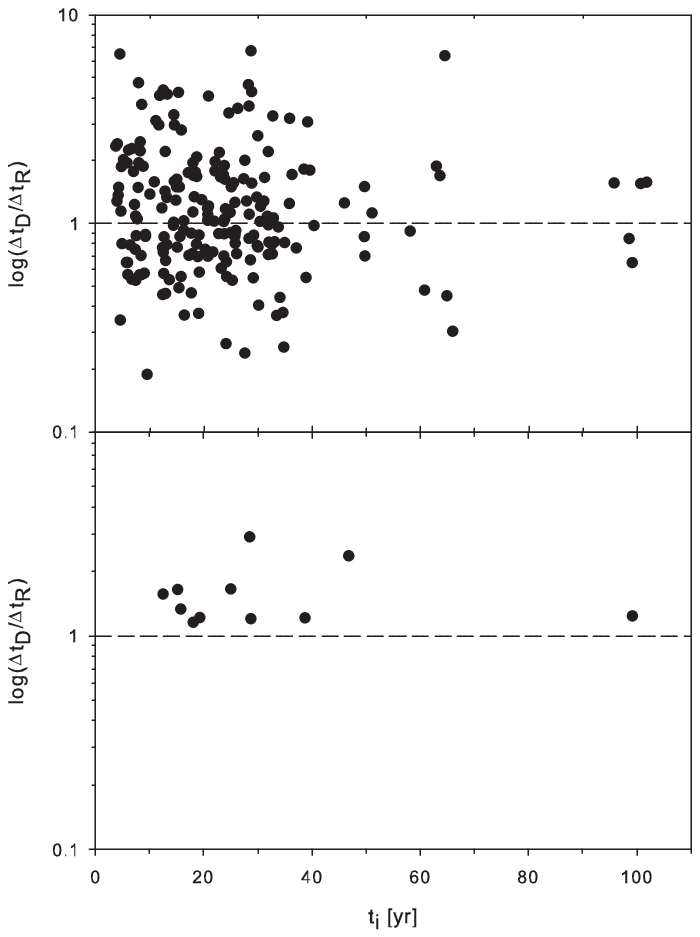}
\caption{The ratio of the decay ($\Delta t_D$) and rise ($\Delta t_R$) times of all flares plotted against the duration of the flares. All values are Doppler-corrected. In the top panel all flares are included and in the bottom panel the source-specific mean values are plotted. For clarity, also the line $\Delta t_D\,/\,\Delta t_R = 1$, where the rise and decay times are of equal length, is included.}
\label{dec_rise}
\end{figure}

In 56\% of all the flares the rise time $\Delta t_R$ is shorter than
the decay time $\Delta t_D$. The ratio of the decay time and rise time
in logarithmic scale is plotted against the duration of the 
flare in Fig.~\ref{dec_rise},
where the values have been corrected for relativistic boosting
according to Eq.~\ref{dopdur}. In the top panel all flares in all
frequencies are included, and in the bottom panel only the
source-specific mean values are plotted. In the top panel, the
individual flares cover the available parameter space quite
well. There are many flares in which the decay time is several times
longer than the rise time and the mean ratio for all the flares 
is 1.61. The mean 
ratio for flares with decay time longer than the rise time is 2.32 while 
for sources with decay time shorter than the rise time the mean ratio is 
0.67.
In the case of the the source-specific mean
values, however, the differences are more moderate, the average ratio
being 1.59. This value corresponds resonably well to the value 1.3 
used by \citet{valtaoja99_tfdI} in the exponential decomposition of radio
flares. The difference between the values can be explained by different 
flare definition. In \citet{valtaoja99_tfdI} individual shock components 
are used while in our approach an activity phase (which may include several 
shocks) is considered as a flare. On average, the decay times are longer 
than the rise times of the flares for all sources. However, as the top 
panel of Fig.~\ref{dec_rise} shows, one source can have flares of very diverse
characteristics.

\placefigure{dec_rise}

\subsection{Time lags of the flare peaks}\label{lag}

We made a qualitative analysis of the individual flares of our sample
sources, tracking the order in which the flare moved from frequency
band to the next and tracing the evolution of the peak flux from its
maximum value. We used all available frequencies for each source. In
flare 1 of 1308+326 and flare 3 of 3C 446, we took into account only
the first component of the flare, although in some frequency bands
other components may be stronger. Of the 45 flares included in our
sample, 11 (24\%) were consistent with the description of a
high-peaking flare: the high frequencies peak first with the highest
peak fluxes. There were nine (20\%) more that were very nearly
consistent, for example, with one frequency band peaking "too
early". In four (9\%) flares we possibly detected also the plateau
stage in the high radio bands after which the flare turns into
high-peaking. There was only one (2\%) flare that was consistent with
the low-peaking flare behavior (flare 9 of OJ 287, defined only at
three frequency bands), and one that was nearly consistent (flare 4 of
BL Lac). Four flares were entirely inconsistent with the shock model,
having no sensible order in either the frequency or flux
evolution. However, all these four flares occurred in sources with
very fast and frequent variability (S5 0716+714, OJ 287, PKS 1749+096
and BL Lac), which means that the different flare components are
particularly hard to separate. It is possible that this has affected
our analysis. For a third of our flare sample, 15 flares in all, we
could make no meaningful analysis of the time lags at all because of 
the sparse sampling of the flares.

We also calculated the time lags for 27 flares with sufficiently
resolved structure. For some of them, time lags could be determined
for only 2 or 3 frequency bands. The mean values of these tentative
time delays range from roughly 10 days to 130 days in the observer's 
frame at 37 and 4.8 GHz, respectively. The mean and median values of 
errors in the peak times, in turn,
are of the order of 28 and 10 days, respectively, for the whole
sample. Thus, the precise time lags cannot be determined, but we chose
rather to examine the sequence in which the flare peak reaches each
frequency. Each frequency band was assigned a rank number. The first
frequency band which displayed the flare was ranked 1, the second was
ranked 2 and so on.

\begin{figure}
\plotone{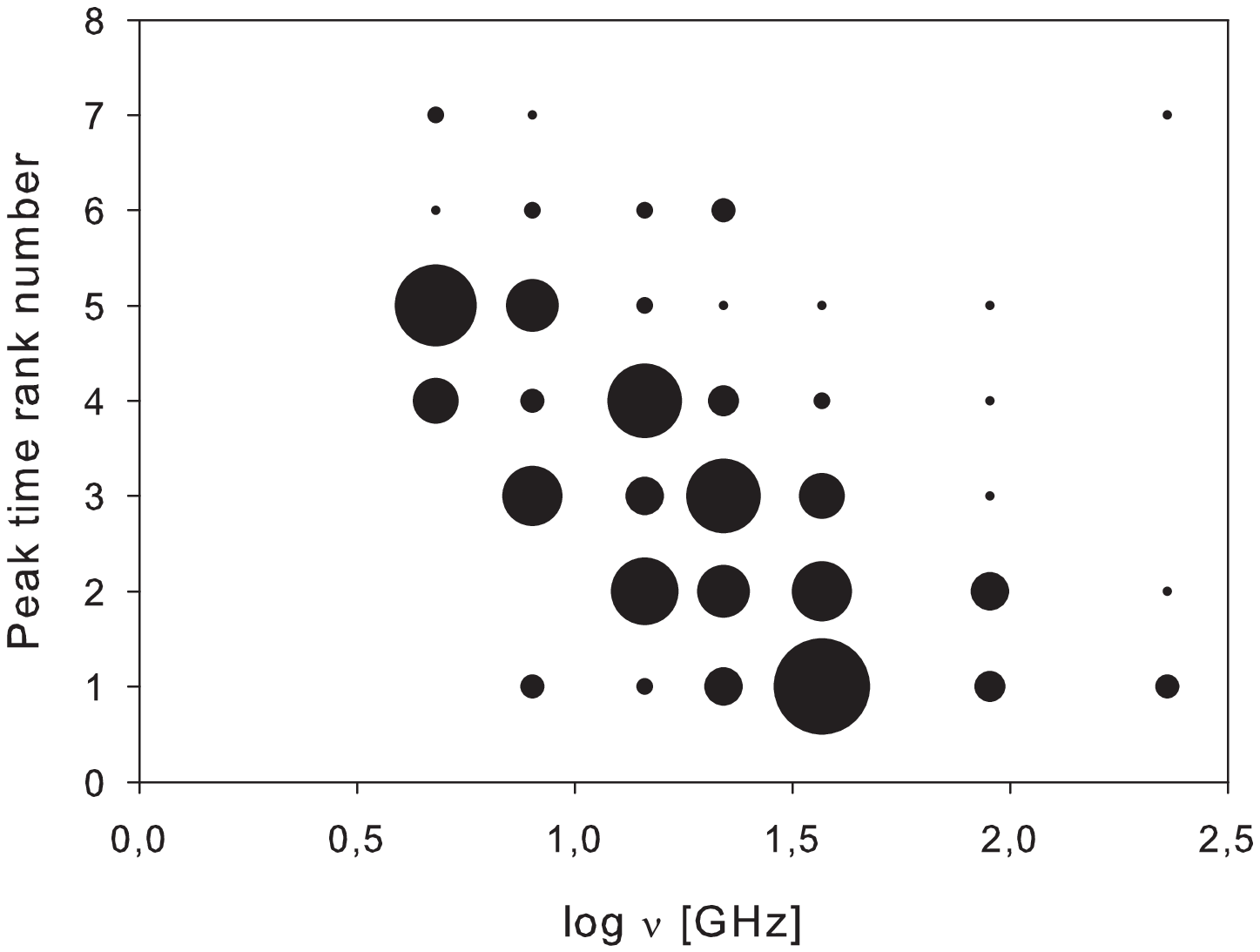}
\caption{Peak time rank number plotted against frequency $\nu$. The frequency band peaking first has been ranked 1.}
\label{bubble}
\end{figure}

\placefigure{bubble}

Figure~\ref{bubble} shows a bubble plot of the peak time rank number
plotted against frequency band. The size of each bubble is
proportional to the number of cases having the same values of rank
number and frequency. According to the general shock model of
\citet{valtaoja92_moniIII}, in high-peaking flares, the higher
frequencies peak first with the low frequencies following with
increasing time delays. That is, if the flares followed the shock
model precisely, we would see a negative correlation in
Fig.~\ref{bubble}, with the highest frequency peaking first (and
having the lowest rank number) and other frequencies following in
order. In Fig.~\ref{bubble} there certainly is a negative trend and
the high frequencies have a lower rank numbers on average. A
significant negative correlation is verified by the Spearman rank
correlation test ($\rho$=-0.567 and $P<$0.0005). In about half of the
flares included in this analysis, the highest frequency band was the
leading frequency band.

\begin{figure}
\plotone{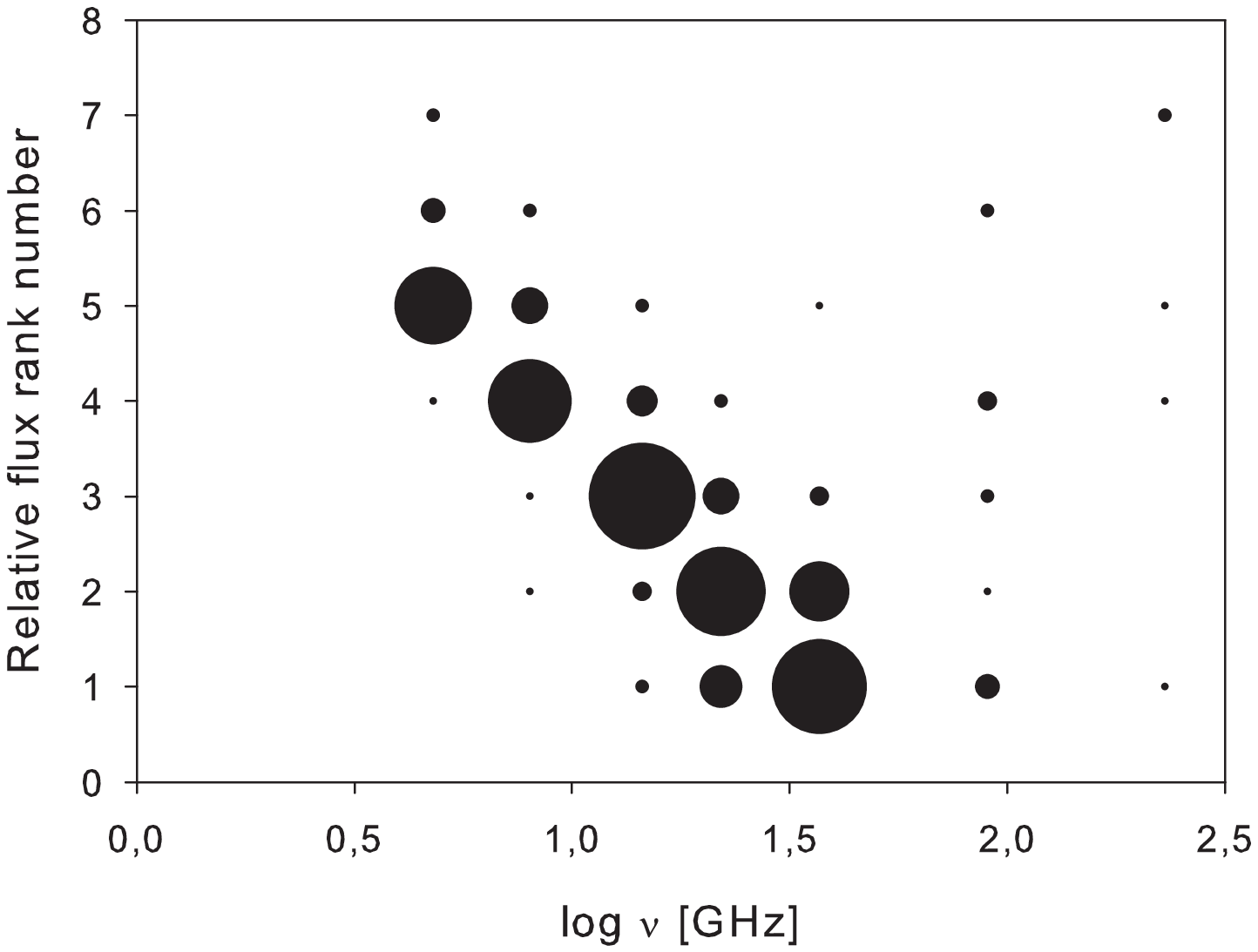}
\caption{Relative peak flux rank number plotted against frequency. The frequency band having the highest relative peak flux has been ranked 1.}
\label{bubble2}
\end{figure}

\placefigure{bubble2}

We plotted another bubble plot (Fig.~\ref{bubble2}) describing the
dependency between the rank orders of relative peak fluxes 
($S_\mathrm{max}-S_\mathrm{min}$) and the
frequency. Of the relative fluxes, the highest was ranked
first. According to the shock model, there should be again a negative
correlation, which is indeed very strong ($\rho$=--0.592 and $P<$0.0005
in the Spearman test). In the two highest frequency bands, 90 and 230
GHz, there are some stray datapoints in the high ranks. There is a strong
possibility that the flares have not been detected in these
frequencies in their full strength due to sparse sampling. The 
plot is very similar when the ratio of maximum and minimum fluxes is used as 
the relative flux.

One should, however, bear in mind that the unambiguous definition of
the time lag of the flare peak from one frequency band to another is
made difficult by the complex structure of some of the flares. When there are
several flare components superposed on each other, it can be tricky to
trace the evolution of just one of them. Also, the error range in the
peak time can be substantial, depending on the sampling density.

\section{The correspondence between time scales and flare parameters}\label{corr}

\subsection{Notes on individual sources}\label{sources}

In the following, we present a brief description of the flaring
 sources and their behavior at all available frequency bands
 individually (flux curves of the sources are available in the
 electronic edition of the Journal). We also compare their flux curves
 to the time scales discussed in \S~\ref{ts}.  \\

\textbf{\textit{S2 0109+22}}: Three distinct flares can be discerned
at 4.8, 8, 14.5, 22 and 37 GHz; the higher frequencies are too
sparsely sampled. The first two flares, peaking in 1993 and 1998,
lasted for about 3 years, while the last one, in 2000, lasted for up
to 6 years in the low frequencies. All three were relatively weak; the
1998 flare had the highest peak flux of 3.13 Jy at 37 GHz. The third
flare reached its peak in approximately one year, depending on the
frequency, but took up to 5.6 years to decrease back to base level
flux. The flares consist of multiple components, which makes it
difficult to find the peak especially in the low frequency flux
curves. In the first flare the 8 GHz flux peaks first, higher
frequencies follow within a month. The 4.8 GHz flux peaks almost 6
months later. In the second one, 37 GHz leads, with the other
frequencies peaking almost a year later, 4.8 GHz being the last again
with a time lag of over a year.

The $T_{SF}$ values for S2 0109+22 range from 0.86 (22 GHz) to 4.817
(8 GHz) yrs. In 22 GHz the flares are much stronger, and the SF is
affected by the relatively well-sampled and fast rises and falls of
the flares. In 8 GHz, the flux curve is less dramatic, and this can be
seen as a longer $T_{SF}$. In 8, 14.5 and 22 GHz the DCF seems to pick
up the interval between flares 1 and 3, giving time scales between
5.82 and 7.73 yrs. At 4.8 GHz, where the peaks of the flares are
barely discernible, $T_{DCF}$= 1.85 yrs. At 37 GHz, the three flares
are clearly above the base level flux, and more evenly separated. This
affects also the DCF, which now gives a time scale of 2.94 yrs, a
little more than half of the interval between flares 1 and 2, and
almost exactly the interval between flares 2 and 3.  \\

\textbf{\textit{AO 0235+164}}: Here we have four distinguishable
multifrequency flares in 1987, 1990, 1992 and 1998. The maximum peak
fluxes are 4.44 Jy at 14.5 GHz, 4.20 Jy at 90 GHz, 6.88 Jy at 37 GHz
and 5.56 Jy at 37 GHz, respectively. The flares have some
substructure, but the peaks are still quite clearly defined. Although
relatively intense, the flares last for 3 years or less in all
frequencies. We were able to determine time lags for two flares. In
both, the flare commences at 230 GHz. The other frequencies follow
some tens of days later in a rather random order. The longest time
delay, 104 days, is at 8 GHz in the 1987 flare, where 230 GHz was the
leading frequency band.
 
The high level of activity in the total flux density flux curve for 
AO 0235+164 is reflected in its SF time
scales, which are mostly below 1 yr. Only at 37 GHz, we find the most 
significant timescale to be $T_{SF}$=\,2.71 yrs, but another 
timescale of $T_{SF}$=\,0.87 yrs is also seen in the SF. 
The approximate intervals between flares 1--\,4
are 4, 2 and 6 years. We find roughly the same numbers listed in
Table~\ref{timescales} as $T_{DCF}$ and $T_{P}$. In most cases, the
most significant time scale is around 5.5 years. In 4.8 GHz, the
periodogram picks up an 11.5 -year time scale, which is twice the
$T_{P}$ of the other frequencies (see also \S~\ref{ts}).  \\

\textbf{\textit{PKS 0422+004}}: The flux curves are quite poorly
sampled, but two events can be discerned. One flare took place in 1994
with a peak flux of 1.37 Jy and another in late 2001 with a peak flux
of 2.40 Jy. These low flux levels make every small variation stand
out, hindering the clear definition of the flare. In the first the
duration ranges from 2.3 to 6 years, depending on frequency band,
while the second one lasted for about 2 years. In the second flare
there are no long time lags, all available frequencies peak within 37
days.

The $T_{SF}$ obtained for PKS 0422+004 are mostly lower limits of
approximately 10 years. For 14.5 and 37 GHz we got $T_{SF}$=\,3.83 and
$T_{SF}$=\,2.71, respectively. The DCF did not produce any time scales
at 22 and 37 GHz, and there were no time scales in the periodogram
analyses of any of the frequency bands. In the lower frequencies,
$T_{DCF}$\,=6--7 years, which corresponds well to the interval of
flares 1 and 2, which is about 7 years.  \\

\textbf{\textit{S5 0716+714}}: This source also has two multifrequency
flares, in late 1998 and 2003. Both flares were monitored in radio
frequencies up to 37 GHz, unfortunately high frequency data is
missing. The first one peaked at 2.54 Jy, while the second one was
considerably stronger at 6.28 Jy. Both peak fluxes occurred at 37
GHz. The latter flare was extensively monitored in a WEBT
multifrequency campaign, including INTEGRAL \citep{ostorero06}. It was
very fast, lasting approximately for 2.5 years. It is noteworthy that
the absolute peak flux of the 2003 flare is strongly dependent on
frequency. At 4.8 GHz it is only 1.9 Jy, and the flare barely stand
out from the base level flux. From there on it steadily rises at each
frequency to exceed 6 Jy at 37 GHz. The same effect can be seen in the
1998 flare, albeit to a lesser extent.

The SF time scales of S5 0716+714 seem to be either very fast
($T_{SF}\leq$1 for 4.8, 14.5 and 37 GHz) or very long ($T_{SF}\geq$\,6
for 8, 22 and 90 GHz). This discrepancy is most likely due to the
sparser sampling of the prominent 2003 flare in 8, 22 and 90
GHz. Especially at 37 GHz it is very frequently sampled, and thus
dominates the flux curve and time scales. The DCF and periodogram time
scales are close to 5 years in 4.8 and 8 GHz, and roughly 2 years in
the higher frequencies. The interval between flares 1 and 2 is of the
order of 5 years. Thus, the two flares seem to define the time scales
in the low frequencies, where they are very low in amplitude, but fail
to do so from 14.5 GHz upward, where they are very strong. This is
probably because of the increasing dominance of the 2003 flare, while
at the same time the sampling in the beginning of the flux curve gets
poorer in the higher frequencies.  \\

\textbf{\textit{PKS 0735+178}}: The flux curve is dominated by a
single, double-peaked flare in all available frequencies. Frequent
sampling and reasonably low short-term variability make its definition
simple. The first component of the double peak was the stronger one in
all frequencies except 4.8 GHz. The peak occurred in most frequencies
in 1989, the peak flux was 5.30 Jy at 14.5 GHz. The duration of the
flare was notably long, ranging from 10 to 12 years. The flux decline
in particular was slow, lasting approximately 8 years.

The SF time scales, determined for frequencies up to 90 GHz, are
diverse, ranging between 2.15 and 6.07 yrs. They are quite long, and
the reason is evident in the flux curves: PKS 0735+178 has minimal
short term flux variation. $T_{P}$ is available for only one
frequency, 8 GHz, being 14.11 yrs. This clearly reflects the modest
pace of the fluctuations of the flux curve. As the source only has one
flare, it is impossible to comment on the compatibility of the flare
intervals and time scales. The interval between the components of the
double peak of flare 1 in of the order of 1.5 years at all
frequencies, distinctly less than any time scale obtained in this
analysis.  \\

\textbf{\textit{PKS 0754+100}}: This source exhibits two modest flares
that have multifrequency data: one peaking in late 1996 and one in
2003. The peaks can be discerned quite effortlessly, although the flux
levels are not very high. The first flare peaks at 2.94 Jy at 37 GHz
and the latter at 2.57 Jy at 14.5 GHz. There was a more intense flare
in the mid-1980's, peaking over 3 Jy at 8 and 14.5 GHz, but it was not
monitored in Mets\"ahovi and thus is not included in our analysis.

The $T_{SF}$, $T_{DCF}$ and $T_{P}$ values could be determined for all
frequencies up to 37 GHz. They are not very consistent: $T_{SF}$ is
approximately 0.5 -- 4.3 yrs, $T_{DCF}$ and $T_{P}$ both approximately
2.8 -- 10.8 yrs. There is also a clear discrepancy between the low and
high frequencies, the latter having much shorter time scales. Much of
the inconsistencies are due to the 1980's flare which is included in
the 4.8 -- 14.5 GHz data, but missing from 22 and 37 GHz data. The
flare was strong and broad, and thus it has lengthened the low
frequency time scales. The interval between the two flares accounted
for in this analysis is approximately 6.5 years at all frequencies,
best reflected by the 22 GHz time scales. The longest time scale,
$T_{DCF}$\,= 10.88 yrs is found at 14.5 GHz. In that frequency band,
the peak of flare 1 is particularly well sampled and strong, probably
strengthening the time scale corresponding to the interval between the
1980's flare and flare 1, which is roughly 12 years.  \\

\textbf{\textit{OJ 287}}: Being one of the most-studied sources in the
Tuorla-Mets\"ahovi observing project, OJ 287 has ample data. Its flux
curves in all frequencies are characterized by vigorous variability
superposed on a long-term fluctuation of the base level flux. The
yearly mean values of flux density at 37 GHz range between almost 8 Jy
and less than 2 Jy. The determination of single flares is very
difficult due to the sheer number of them. We count as many as 9
multifrequency outbursts during the 25 years of our monitoring. They
are quite brief in duration, typically lasting for less than two
years, or, in many cases, less than a year. The highest peak flux,
9.63 Jy, occurred in 1983 at 22 GHz. The undertakings of OJ 287 are of
special interest because of its claimed optical periodicity. This
almost 12-year periodicity is thought to be possibly the result of a
binary black hole interaction \citep{sillanpaa88, valtonen96,
valtonen08}.

The rapid and frequent outbursts result in very short SF time scales,
between 0.2 and 0.5 years at all frequencies. The DCF time scales,
however, determined for 8 and 37 -- 230 GHz, are considerably
longer. They range approximately between 4 and 6.5 yrs. The DCF
probably picks up also the base fluctuation, which can have a time
scale as long as 20 years. The sole $T_{P}$ value that could be
calculated was 1.03 yrs at 90 GHz. It describes the source well: the
calculation of the flare intervals gives mostly values below 2
years. The intervals between flares 1 and 2 as well as 6 and 7 are of
the order of 5 years. There are, however, minor flares evident in the
flux curves during those intervals as well, but they are not included
in this analysis.  \\

\textbf{\textit{1308+326}}: This source had a flare that lasted for
the entire 1990's. It peaked in 1992 at 4.37 Jy at 22 GHz. In the
lower frequencies it was clearly double-peaked, and had several
components also in 22 and 37 GHz. After a slow decline the flux levels
soared again in 2003, reaching a peak flux of 3.5 Jy at 14.5 GHz. An
optical period for the source has been reported \citep{fan02}, but no
periodicity in the radio data has been detected to our knowledge.

The interval between the two flares of 1308+326 is 8 years at 8 GHz
and roughly 11 years at 14.5 -- 37 GHz. This is evident in the
$T_{DCF}$ --values, which are between 10.34 and 11.70
yrs. \citet{pyatunina07} derived an activity cycle of $\geq 14$ yrs
for this source, which also reflects its long time scales of
variability. At 90 GHz, $T_{DCF}$\,= 3.77 yrs, but this value has
exceptionally large errors and is based on a poorly sampled flux curve
where the flares cannot be discerned properly. The periodogram gives
mostly similar results as DCF. At 8 GHz $T_{P}$\,= 14.59 yrs, which is
quite high compared to the $T_{DCF}$ --values. The SF time scales
range as $T_{SF}$\,= 2.71 -- 4.29, telling that 1308+326 has little
short term variability.  \\

\textbf{\textit{PKS 1413+135}}: The flux curves are very erratic,
making it difficult to find well-defined outbursts. Two multifrequency events 
can be discerned. The first flare peaks at 4.55 Jy at 37 GHz in the end of
the year 1990. The second peaks at the same frequency at 2.46 Jy 7
years later, in 1997. The peak of the latter flare in particular is
ill-defined. There is a clear flux boost lasting for more than six
years, during which there are several minor maxima in all
frequencies. There is also a pronounced flare visible in the low
frequency data in the early 1980's, but the high frequency data is
missing.

Noteworthy is also the strong evolution of the flux curve with
frequency. At 4.8 GHz the flux curve is all but flat, but gradually it
gets more eventful towards higher frequencies. This is reflected by
the time scales. Both $T_{SF}$ and $T_{DCF}$ are clearly longer in the
low frequencies. $T_{SF}$ is 2.71 yrs at 8 GHz and 1.52 at 37 GHz;
$T_{DCF}$ is 9.38 yrs at 4.8 and 8 GHz, declining slowly to 2.26 at 90
GHz. Curiously enough, $T_{P}$ values seem to remain unaffected by the
evolution of the flux curve with frequency, ranging between roughly 7
and 9 years. In addition, the 14.5 GHz time scales are affected by the
1980's flare which is not included in the high frequency data. The
intervals of the flares are hard to determine because of the ambiguous
definition of the peak of the second flare. At 8 GHz it is roughly 4
years, at 14.5 GHz 2 years and at 22 and 37 GHz roughly 7 years. Thus,
it is approximately comparable to $T_{P}$, except at 14.5 GHz.  \\

\textbf{\textit{PKS 1749+096}}: This object exhibits violent
variability. There are five multifrequency 
flares peaking in 1993, 1995, 1998, 2001
and 2002. The 1993 flare had the highest peak recorded in our data,
12.07 Jy at 90 GHz. Given the relative brevity of the flares and their
intensity, PKS 1749+096 also exhibits some of the fastest grand-scale
rises and declines of flux. For example, in the 1993 flare it reached
the peak flux in approximately 80 days, rising over 9 Jy. While
normally a flat spectrum source, as BLOs in general, this object has
been reported to have an inverted spectrum during outbursts
\citep{torniainen05}.

PKS 1749+096 is another example of a flux curve evolving with increasing
frequency. The flares, barely discernible at 4.8 GHz, get more intense
at higher frequencies. As in the case of PKS 1413+135, the evolution
is reflected in the time scales, albeit less clearly. Mostly it can be
seen in the $T_{DCF}$ values. They decline from 4.45 yrs at 4.8 GHz to
1.30 yrs at 90 GHz. At 8 GHz, $T_{DCF}$ is the longest, 6.78 yrs. The
intervals between the flares 1 -- 4 are roughly 2 -- 3 years. The
interval between flares 4 and 5 is approximately 1 year. Also, as with
PKS 1413+135, the $T_{P}$ seem to be less affected by the changing
behavior of the flux curve with growing frequency. $T_{P}$ has its
highest value at 37 GHz, where $T_{P}$\,= 9.81. The SF time scales are
quite short, below 2 years, except for 8 GHz ($T_{SF}$\,= 2.15) and 90
GHz ($T_{SF}$\,= 3.04). The beginning of the flux curve is
particularly well-sampled and relatively stable at 8 GHz, which
probably lengthens $T_{SF}$ compared to the high frequencies, which
have less data. On the whole, the time scales of 14.5 and 22 GHz seem
to correspond best to the flares included in the analysis.  \\

\textbf{\textit{S5 2007+77}}: This BLO has one outburst, which was
reasonably well-monitored also at the Mets\"ahovi frequencies. It
occurred in 1991-1992 and reached a peak flux of 3.69 Jy at 14.5
GHz. The low frequency flux curves reveal that the source was very
variable also prior to that time, but since the early 1990's it has
been in a quiescent state with only modest variability. Unfortunately,
the flux curves are not very well-sampled and data from 8 GHz and the
very highest frequencies, 90 and 230 GHz, are missing completely from
the flare analysis.

The SF time scales are mostly the typical 0.5 -- 2 yrs, only at 22 GHz
it is as long as 3.83 yrs. The 22 GHz flux curve, however is
under-sampled and the time scales are tentative at best. In 37 GHz,
the $T_{SF}$ did not show a plateau and no time scale could be
determined. The rise and decay times of flare 1 are between 1 and 2
years, so $T_{SF}$ at 4.8 and 14.5 GHz describe them
accurately. $T_{DCF}$ produced significant time scales only at 4.8, 8
and 37 GHz, and $T_{P}$ only at 37 GHz. All values are approximately 2
-- 3 years, roughly compatible with the 1980's flaring evident in the
low frequency flux curves.  \\

\textbf{\textit{BL Lac}}: The archetype of all BL Lacertae objects
certainly has a very variable flux curve. Unfortunately the major
flare of early 1980's is not included in our analysis due to poor
sampling in 22 and 37 GHz. Since then, however, we count nine
multifrequency 
outbursts in 1987-1988, 1989, 1992, 1993, 1996, 1997-1998, 2000,
2002 and 2003. All are similar in shape in the higher frequencies,
peaking at 3--6 Jy. At 4.8 and 8 GHz the 1987-1988 and 1996 flares
(flares 1 and 5) are more prominent with fast rises, and outbursts
following them seem to be partially superposed on their slow
decline. With growing frequency, flares 1 and 5 seem to lose their
dominance and blend into the other flares.

$T_{SF}$ values are mostly below 1 year, except for 4.8 GHz
($T_{SF}$\,= 3.83) and 22 GHz ($T_{SF}$\,= 2.41). In the former case,
flares 1 and 5 dominate the small scale variations, leading to a long
SF time scale. Especially the decay time of flare 1 is long. In 22
GHz, the reason for a long SF time scale is less clear. In the case of
$T_{DCF}$ and $T_{P}$, long time scales clearly dominate. The typical
values are $T_{DCF}$\,= 7.5 yrs and $T_{P}$\,= 8.5 yrs. This contrasts
with the average flare interval of approximately 2 yrs. The reason for
such long time scales is probably the strong flare on early 1980's,
which is unequalled in flux density. This conclusion can also be drawn
from the 8 GHz time scales. They are considerably shorter,
$T_{DCF}$\,= 2.81 yrs and $T_{P}$\,= 3.80 yrs, while 8 GHz is also the
only frequency that has data from the beginning of 1970's when BL Lac
was very bright, flux levels being comparable to those of the 1980's
flare.  \\

\textbf{\textit{3C 446}}: The flux curve is marked by three flares,
two of them quite broad and strong. The first peaked approximately in
1990. In 230 GHz, however, the peak occurred as early as 1988, when
the flux levels reached 11.71 Jy. The other two peaked in 1996 and
2000, at 6.30 Jy and 9.29 Jy, respectively, at 22 GHz. All outbursts
have multiple components and frequency evolution is apparent.

3C 446 has relatively few short scale flux variations and rise and
decay times of 3 to 4 yrs in flares 1 and 3. This can be seen in the
SF time scales which are quite long, roughly $T_{SF}$\,= 3 yrs, except
for 8 GHz, for which $T_{SF}$\,= 1.5 yrs. This could be due to either
the randomly sampled early flux curve at 8 GHz, which is missing in
other frequencies, or just minor variations. The DCF and periodogram
time scales were significant only at the low frequencies, and
$T_{DCF}$ also at 90 GHz. All $T_{DCF}$ and $T_{P}$ are long, mostly
roughly 10 years or above, reflecting the rather sedate behavior of
the source. However, their behavior is not very consistent: at 14.5
GHz, $T_{P}$\,= 5.71 yrs which is considerably shorter than at 4.8 and
8 GHz, but $T_{DCF}$\,= 11.16 yrs which is clearly longer than at the
lower frequencies. It is of the same order as the 12 -year activity
cycle for 3C 446 obtained by \citet{pyatunina07}. At 90 GHz,
$T_{DCF}$\,= 5.96 yrs, which roughly corresponds to the 6 -- 7 year
interval between flares 1 and 2. The average interval between flares 2
and 3 is 3.5 -- 5 years.

\subsection{Correlations}

\begin{figure}
\plotone{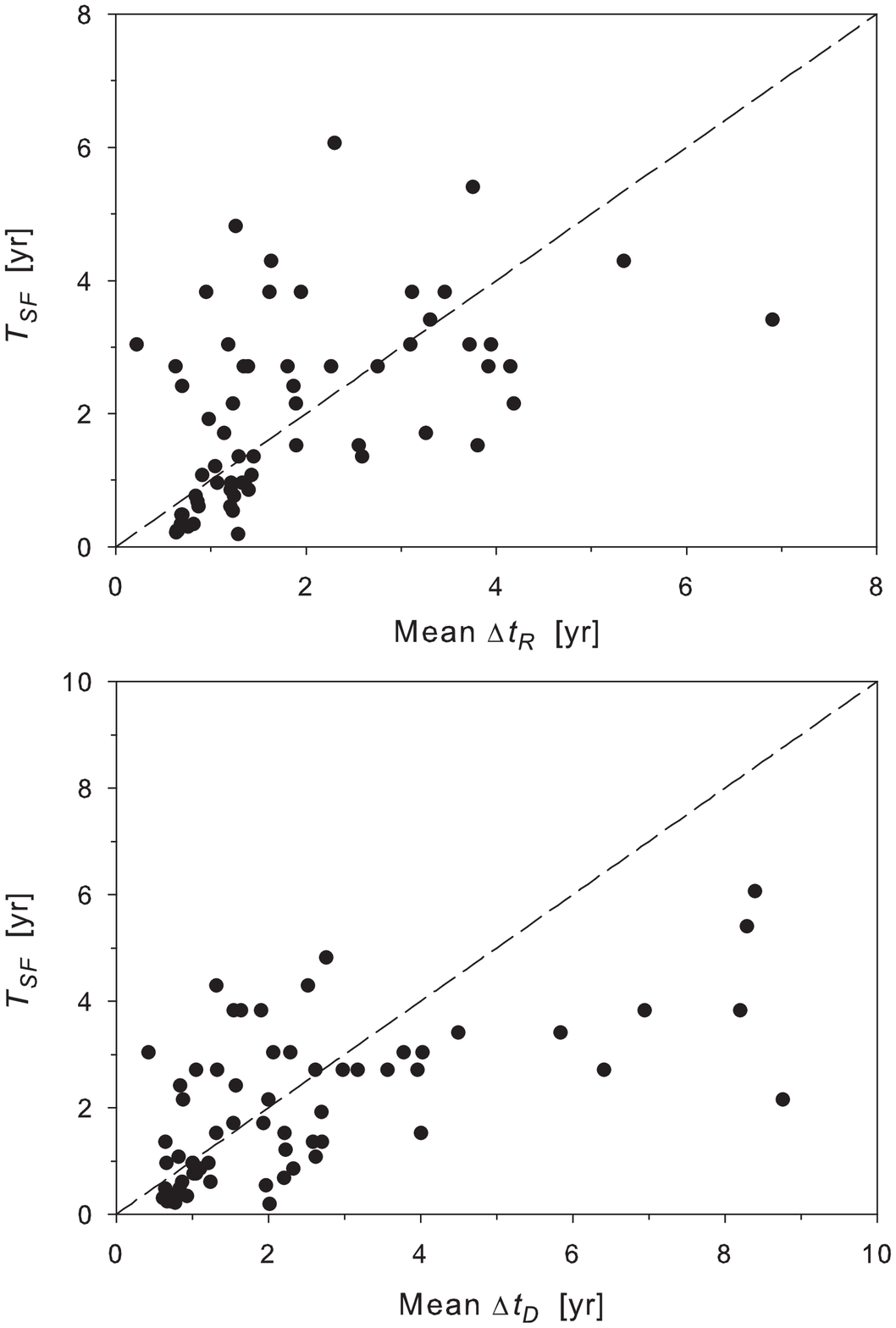}
\caption{$T_{SF}$ plotted against the rise times, $\Delta t_R$ (top panel), and decay times, $\Delta t_D$ (bottom panel), of the flares, averaged for each source and frequency. The dashed line represents a one-to-one correspondence.}
\label{SF_vs_rise_decay}
\end{figure}

\begin{figure}
\plotone{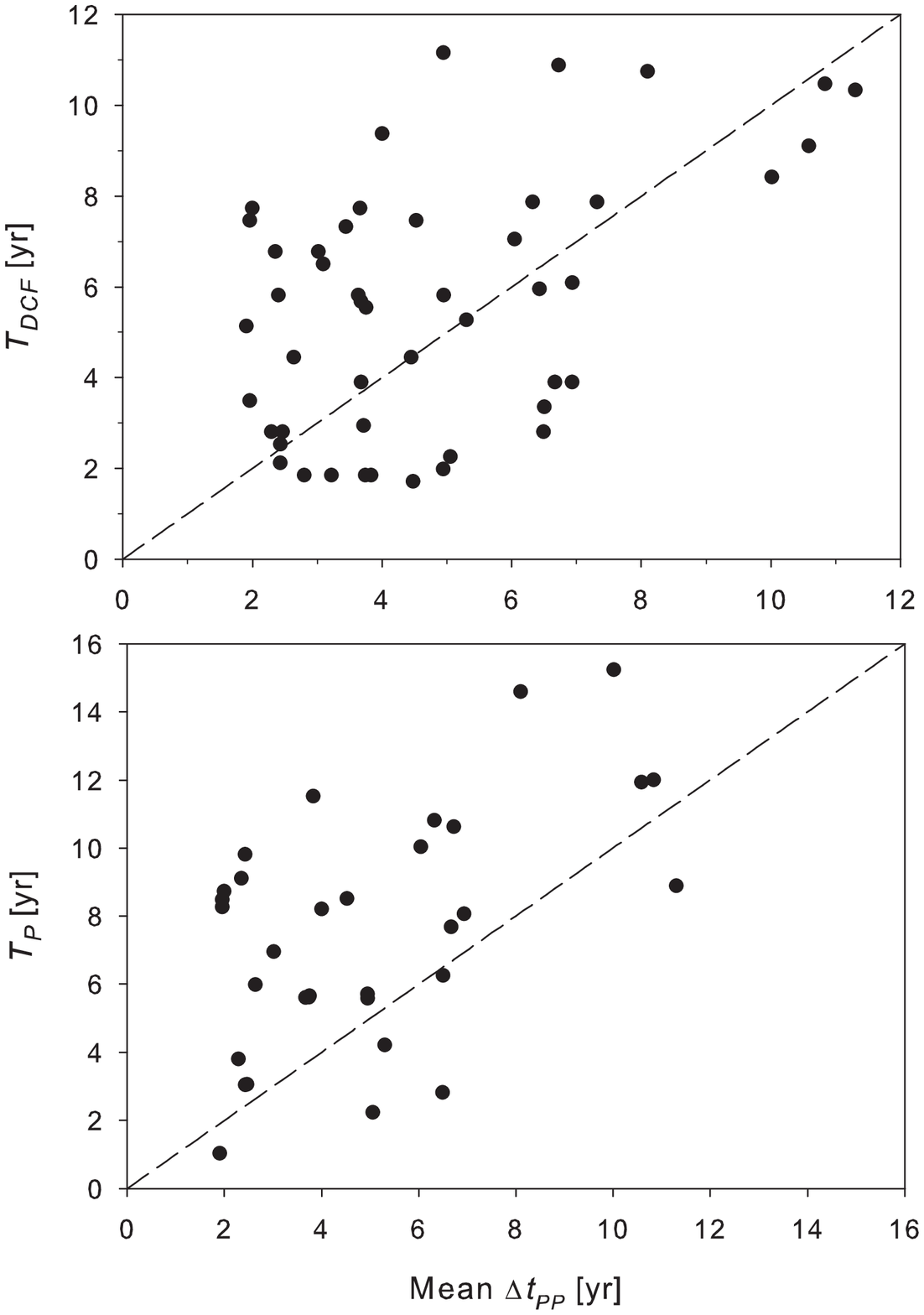}
\caption{$T_{DCF}$ and $T_{P}$ plotted against the peak-to-peak intervals, $\Delta t_{PP}$ (top panel) of the flares, averaged for each source and frequency. The dashed line represents a one-to-one correspondence.}
\label{ptp}
\end{figure}

In \S\ref{sources} we described in detail how, for each source, the
time scales and the observed flux curve related to each other. To
illustrate the correspondence between the time scales and the temporal
parameters of the outbursts more quantitatively we plotted i) SF time
scales of the source against both the rise times, $\Delta t_R$, and
the decay times, $\Delta t_D$, of the flare, averaged for each source
and frequency (Fig.~\ref{SF_vs_rise_decay}), ii) DCF and periodogram
time scales against the peak-to-peak intervals of consecutive flares,
$\Delta t_{PP}$, averaged for each source and frequency
(Fig.~\ref{ptp}). In all the plots, the dashed line represents an
ideal one-to-one correspondence. All parameters are observational and
have not been corrected for redshift nor for Doppler boosting.

\placefigure{SF_vs_rise_decay}

From Fig.~\ref{SF_vs_rise_decay} we see that the plot is very similar
for both the rise and decay times. There is considerable scatter at
low $\Delta t_R$ and $\Delta t_D$ on both sides of the one-to-one
line, and at high values the SF time scales seem to be significantly
shorter than $\Delta t_D$. The lower limits of $T_{SF}$ were not
included in the plot. According to the Spearman rank correlation test,
there is a significant positive correlation between $T_{SF}$ and both
$\Delta t_R$ and $\Delta t_D$. For rise times, $\rho$=0.600 and
$P<$0.0005, and for decay times $\rho$=0.607 and $P<$0.0005.

\placefigure{ptp}

The distribution of the $T_{DCF}$ and $T_{P}$ values plotted against
$\Delta t_{PP}$ (Fig.~\ref{ptp}) are also scattered, but seems to
roughly follow the one-to-one line. According to the Spearman test,
the positive correlation is significant for both the DCF ($\rho$=0.366
and $P$=0.005) and periodogram ($\rho$=0.420 and $P$=0.008).

\section{Discussion}\label{dis}

Throughout this paper it is important to remember that this sample
represents only a small fraction of the BLO population. Most BLOs are
too faint in the radio frequencies, or even if their flux density is
above the detection limit, they simply lack the long-term data needed
for this kind of analysis. Also, the high-energy BLOs (HBLs) are
sorely underrepresented: only two of them are included in the time
scale analysis, and none at all in the flare analysis. It is also
noteworthy that only 13 of the 24 sources included in the time scale
analysis had significant, well-sampled flares to analyze during the
observing period. Some of the remaining 11 sources simply do not have
a very variable flux curve,
which indicates that even some radio-bright BLOs are surprisingly
steady emitters. For example, Mark 421, B2 1147+24, and Mark 501
have remarkably uneventful radio flux curves. This is also confirmed 
by other authors \citep[e.g.][]{venturi01, blazejowski05, lichti08}.

While the shock-in-jet scenario gives the general guidelines of AGN
variability and its causes, there are many additional factors, such as
relativistic boosting, properties of the ambient medium, turbulence
and bending of the jet \citep{marscher96}, affecting the flux behavior
we observe. These effects together with the shock mechanism generate
the diverse flux curves observed also in our sample, ranging from the
rapid spikes of OJ 287 to the modest pace of 1308+326 and PKS
0735+178. 

As stated in Sect.~\ref{lag}, there are two things that complicate the
analysis of the time lags of the BLO flares: the very complex
structure of most of the flares and the regrettably sparse
sampling. The first affects especially the small flares, where it can
be impossible to separate the components from each other and thus
their evolution cannot be traced. The latter creates errors in both
the peak time and peak flux of the flare. In many cases errors in the
peak time in different frequency bands do not allow the unambiguous
determination of the order in which the flare peak reaches each band.

In spite of these complications, we find that most of the BLO flares
seem to be high-peaking in the radio frequencies, following the
classification of \citet{valtaoja92_moniIII}. The possible detection
of the flare peak plateau in $S$ vs. log $\nu$ -representation in some
cases, and the relatively short time delays in the high radio
frequencies lead us to believe that the BLO flares typically reach
their maximum development stage in the mm-or sub-mm-wavelengths. There
were 4 flares, which were inconsistent with the generalized shock
model. This is probably due to errors in peak timing and the
underlying complexity of the flare.

Our view of the high-peaking nature of BLO radio flares is supported
by the general behavior of the multi-frequency flux curves. In many
sources, most notably S2 0109+22, S5 0716+714, OJ 287 and PKS 1413+135
the increasing flare flux levels with increasing frequency are evident
just by looking at the data. The base level fluxes do not rise as
steeply, which suggests that the flux increase can be attributed to
the flaring component, i.e., the shock. There are exceptions; for
example, PKS 0735+178 and 1308+326 do not follow this rule to the same
extent. As we saw in \S\ref{mor}, their behavior stands out in other
ways as well.

Helical / curved jets have also been suggested to be the cause 
of large variations in the flux curves of many BLOs. \citet{villata99} 
developed 
a model to explain the spectral variations of Mark 501 with a helical 
jet produced by a binary black hole system. The model was able to describe 
the peculiar X-ray part 
of the SED very well but the low frequency optical to radio part 
remained fairly constant. They concluded that the low frequency variations 
could be due to inhomogenities in the rotating jet or intrinsic brightness 
variations. \citet{ostorero04} applied the model to the SED and the radio 
and optical flux curves of AO 0235+164. The model was based on the 
5.7 year quasi-periodicity suggested for the source by \citet{raiteri01}. The 
periodic flares are explained by rotation of the helix. Observed signatures
are similar to the shock model so that first the high frequency portion 
of the jet approaches the line of sight and Doppler boosting increases causing
the flux density to rise. As the helix rotates, different frequency 
portions approach the line of sight and this way the time delays between 
the frequency bands can be explained. In their model, the non-periodic 
flares were explained with intrinsic brightness variations (e.g. shocks). 
As the model was based on the observed periodicity of 5.7 years, it 
should be modified now that the period did not repeat after the year 2000 
\citet{raiteri06}. It should be noted that the sources in our sample 
are not strictly periodic in the radio regime and usually the observed 
quasi-periodicities last only a short time in the flux curve 
\citep{hovatta08a}.

The model by \citet{villata99} has also been used to explain 
the variations in BLOs S4 0954+65 \citep{raiteri99}, ON 231 \citep{sobrito01} 
and S5 0716+714 (\citealt{ostorero01}, using data prior to the extreme 
flare in 2003). In addition, VLBI polarization observations have revealed 
helical magnetic fields in many BLOs \citep[e.g.][]{gabuzda04, mahmud08}.
Some of those are also in our sample, but they are mostly sources for 
which we have not performed detailed flare analysis because they do not 
have distinct flares in the radio frequencies. It is indeed possible that 
in these sources the variations are caused by changes in the Doppler beaming 
due to curved jets rather than intrinsic phenomena like shocks. 
However, testing this scenario would require detailed studies of 
simultaneous SEDs, which is beyond the scope of this paper.

We showed in \S\ref{corr} that the computational time scales correlate
fairly well with the observed temporal parameters. To our knowledge,
such a straightforward, but revealing comparison has not been done
before. The statistically significant correlation in
Figs.~\ref{SF_vs_rise_decay} and \ref{ptp} confirms that the SF, DCF
and L-S periodogram time scales are indeed directly linked to the
source behavior we observe. Unfortunately, the scatter is
substantial. For example, in our data a source with a DCF time scale
close to 8 years, can have real peak-to-peak flare interval of 2 to 10
years. The average absolute deviation of the computed time scale from
the one-to-one correspondence is 0.98 and 1.24 years for $T_{SF}$
against $\Delta t_R$ and $\Delta t_D$, respectively, and 2.24 and 3.23
years for $T_{DCF}$ and $T_{P}$ against $\Delta t_{PP}$,
respectively. In representing the peak-to-peak intervals of the
flares, $T_{DCF}$ and $T_{P}$ are near equivalent. However, at least
in the cases of PKS 1413+135 and PKS 1749+096, the periodogram results
are less affected by the frequency evolution of the flux curve between
the frequency bands.

Variability of BLOs in the lower radio frequencies of 4.8, 8 and 14.5
GHz was also studied by \citet{aller99}. They studied the variability
behavior of a complete flux-limited sample of 41 BLOs using e.g. the
SF. Only two of the BLOs in our sample (OJ 425 and 4C 56.27) are not
included in the sample of \citet{aller99}. They used UMRAO data from
1980 to 1996, while we have used the same database updated until April 2005. We
have compared our results to see if the source behavior has changed
during the past ten years. On average, the time scales have remained
quite similar, the average longest time scale from SF in
\citet{aller99} is 2.9 years compared to our average SF time scale for
BLOs which is 3.7 years (median is 2.7 years) \citep{hovatta07}. When
individual sources are studied, there are some differences and we
believe it is mainly due to the longer dataset used in our
analysis. Similar results were obtained in \citet{hovatta07} when the
SF time scales for the whole AGN sample at 22 and 37 GHz were compared
to analyses made ten years earlier.

In many of the BLOs in our sample short intraday variations 
are seen in optical and radio frequencies \citep[e.g.][]{wagner95}. 
However, the sampling 
density of our monitoring programs is not frequent enough to detect 
such rapid variations, unless a special campaign is arranged. The only 
example for which intraday variations have been observed at 37\,GHz using 
Mets\"ahovi data is S5 0716+714 \citep{ostorero06}. During the extreme 
flare in 2003 a rapid flux rise of 42\% was observed in a time period 
of 0.12 days. During the observing campaign the source was observed 
multiple times in a day, while usually our sampling density is of 
the order of a week. \citet{ostorero06} also concluded that the rapid 
variations are intrinsic to the source and not caused by interstellar 
scintillation (ISS), which is a more pronounced phenomenon at low radio 
frequencies. \citet{ricket06} have studied ISS of 146 extragalactic 
radio sources at 2 and 8\,GHz, and their sample includes 13 sources which 
are also in our sample. For these sources the typical ISS timescales range 
from 2.9 to 10.4 days at 2\,GHz and from 4.4 to 18.6 days at 8\,GHz. 
Also, the ISS contribution is much weaker at 8\,GHz than at 2\,GHz. The 
timescales are so short that the variations caused by ISS would mostly 
not be seen in our observations. They also conclude that for some of the 
well-studied BLOs in our sample (e.g. AO 0235+164, PKS 1749+096 and 
BL Lac) it is clear that intrinsic variations dominate. Therefore 
we do not think that our results are affected by ISS.

\section{Conclusions}\label{con}
We have studied the long-term radio variability and the flare morphology of radio-bright BL Lacertae objects. The main conclusions are as follows:

\begin{enumerate} 
\item Radio-bright BLOs exhibit a range of flaring behavior, with few
common features. Especially the quasar-like objects PKS 0735+178,
1308+326 and 3C 446 have distinctively long outbursts with modest
short-term variability, contrasting with the more erratic behavior of
other sample sources. The long flare of 1308+326 clearly stands out
from the rest of the sample even after correcting for the relativistic
boosting effects. Our findings confirm the quasar-like nature of
1308+326 and 3C446 and indicate that PKS 0735+178 also has radio 
behavior different from typical BLOs.

\item The median duration of a flare in a radio-bright BLO is of the
order of 2.5 years, and the peak flux density typically reaches about
5 Jy at 37 GHz. On average, the decay time of the flare is 1.6 times
longer than the rise time. When the Doppler boosting effect is taken
into account, the peak flux of the flare does not depend on the
duration of the flare, indicating that the energy release in a flare
does not depend on its duration.

\item The 45 BLO flares in our analysis confirm the generalized shock
model of \citet{valtaoja92_moniIII} with no clear, undisputed
exceptions. However, very frequent sampling on several radio
frequencies is needed for the accurate, observational determination of
the flare components and time lags. Based on the evolution of the
relative peak flux and the time lags from one frequency band to the
next, we find that the BLO flares are mostly high-peaking. Probably
they reach their maximum development in the mm- to sub-mm-wavelengths.

\item The computational time scales, $T_{SF}$, $T_{DCF}$ and $T_{P}$,
have a statistically significant correlation with the temporal flare
parameters obtained directly from the flux curves. However, scatter is
considerable, and the average deviation from one-to-one correspondence
is of the order of 1 - 3 years, depending on the parameter and time
scale in question.

\end{enumerate}

\acknowledgements
We gratefully acknowledge the funding from the Academy of Finland (project numbers 205793, 210338 and 212656). UMRAO is supported in part by a series of grants from the NSF, most recently AST 0607523, and by funds from the University of Michigan Department of Astronomy.




\begin{deluxetable}{l l l l l l c c l}
\tablecaption{The source sample of 24 BLOs. Column (8) indicates whether the source is included in the flare analysis. See text for details.\label{sample}}
\tablewidth{0pt}
\tabletypesize{\footnotesize}
\tablehead{\colhead{Source} & \colhead{Alias} & \colhead{R.A.(J2000)} & \colhead{Dec(J2000)} & \colhead{z} & \colhead{$D_{var}$} & \colhead{Class} & \colhead{Flare analysis} & \colhead{ref. for z}}
\startdata
0109+224 & \object{S2 0109$+$22} & $01^h12^m05.8^s$ & $+22^\circ44'39''$ & \nodata  &  \nodata & LBL & * & \nodata\\
0235+164 & \object{AO 0235$+$164} & $02^h38^m38.8^s$ & $+16^\circ36'59''$ & 0.940  &  24.0 & LBL & * & 3 \\
0422+004 & \object{PKS 0422$+$004} & $04^h24^m46.8^s$ & $+00^\circ36'07''$ & 0.310  &  \nodata  & IBL & * & 9 \\
0716+714 & \object{S5 0716$+$714} & $07^h21^m53.3^s$ & $+71^\circ20'6''$ & 0.310   & 10.9  & LBL & * & 8 \\
0735+178 & \object{PKS 0735$+$178} & $07^h38^m07.4^s$ & $+17^\circ42'19''$ & 0.424  & 3.8  & LBL & * & 5 \\
0754+100 & \object{PKS 0754$+$100} & $07^h57^m06.7^s$ & $+09^\circ56'35''$ & 0.266  & 5.6 &  LBL & * & 2 \\
0814+425 & \object{OJ 425} & $08^h18^m16.1^s$ & $+42^\circ22'46''$ & 0.245  & 4.6 & LBL & & 1 \\
0851+202 & \object{OJ 287} & $08^h54^m48.8^s$ & $+20^\circ06'30''$ & 0.306  &  17.0 & LBL & * & 12 \\
0954+658 & \object{S4 0954$+$65} & $09^h58^m47.2^s$ & $+65^\circ33'54''$ & 0.367  & 6.2  & LBL & & 7 \\
1101+384 & \object{Mark 421} & $11^h04^m27.2^s$ & $+38^\circ12'32''$ & 0.031  & \nodata &   HBL & & 16 \\
1147+245 & \object{B2 1147$+$24} & $11^h50^m19.2^s$ & $+24^\circ17'54''$ & 0.200  & \nodata  & LBL & & 10 \\
1219+285 & \object{ON 231} & $12^h21^m31.7^s$ & $+28^\circ13'58''$ & 0.102  &  1.2  & IBL & & 18 \\
1308+326 & \object{AUCVn} & $13^h10^m28.7^s$ & $+32^\circ30'43.8''$ & 0.997  & 15.4  & LBL & * & 15 \\
1413+135 & \object{PKS 1413$+$135} & $14^h15^m58.8^s$ & $+13^\circ20'24''$ & 0.247  &  12.2 &  LBL & * & 19 \\
1418+546 & \object{OQ 530} & $14^h19^m46.6^s$ & $+54^\circ23'14''$ & 0.151  & 5.1 & LBL & & 13 \\
1538+149 & \object{4C 14.60} & $15^h40^m46.5^s$ & $+14^\circ47'45.9''$ & 0.605  &4.3 & IBL & & 14 \\
1652+398 & \object{Mark 501} & $16^h53^m52.2^s$ & $+39^\circ45'36''$ & 0.034  & \nodata & HBL & & 16\\
1749+096 & \object{OT 081} & $17^h51^m32.7^s$ & $+09^\circ39'01''$ & 0.322  & 12.0 & LBL & * & 11 \\
1803+784 & \object{S5 1803$+$784} & $18^h00^m45.4^s$ & $+78^\circ28'04''$ & 0.684  &  12.2 & LBL & & 6 \\
1807+698 & \object{3C 371.0} & $18^h06^m50.7^s$ & $+69^\circ49'28''$ & 0.051  & 1.1 & IBL & & 4 \\
1823+568 & \object{4C 56.27} & $18^h24^m07.1^s$ & $+56^\circ51'01.5''$ & 0.663  &  6.4 & LBL & & 7 \\
2007+776 & \object{S5 2007$+$77} & $20^h05^m31.1^s$ & $+77^\circ52'43''$ & 0.342  & 7.9 & LBL & * & 13 \\
2200+420 & \object{BL Lac} & $22^h02^m43.3^s$ & $+42^\circ16'39''$ & 0.069  & 7.3 & LBL & *& 17 \\
2223-052 & \object{3C 446} & $22^h25^m45.1^s$ & $-04^\circ56'34''$ & 1.404  & 16.0 & LBL & * & 20\\
\enddata
\tablerefs{(1) \citet{britzen08}; (2) \citet{carangelo03}; (3) \citet{cohen87}; (4) \citet{deGrijp92}; (5) \citet{hewitt87}; (6) \citet{hewitt89}; (7) \citet{lawrence86}; (8) \citet{nilsson08}; (9) \citet{smith95}; (10) \citet{sowards05}; (11) \citet{stickel88}; (12) \citet{stickel89}; (13) \citet{stickel91}; (14) \citet{stickel93}; (15) \citet{tytler92}; (16) \citet{ulrich75}; (17) \citet{vermeulen95}; (18) \citet{weistrop84}; (19) \citet{wiklind97}; (20) \citet{wright83}}
\end{deluxetable}


\begin{deluxetable}{llllllc}
\tablecaption{\label{timescales}The most significant timescales obtained for the sample sources using the structure function ($T_{SF}$), the discrete correlation function ($T_{DCF}$) and the Lomb-Scargle periodogram ($T_{P}$). All time scales are in the observer's frame.}
\tablewidth{0pt}
\tabletypesize{\small}
\tablehead{\colhead{Source} & \colhead{Alias} & \colhead{$\nu$ [GHz]} & \colhead{$T_{SF}$ [yr]} & \colhead{$T_{DCF}$ [yr]} & \colhead{$T_{P}$ [yr]} & \colhead{Note}}
\startdata
0109+224 & S2 0109+22 & 4.8 & 1.358 & 1.848 & \nodata & 1 \\
0109+224 & S2 0109+22 & 8 & 4.817 & 7.324 & \nodata & 1 \\
0109+224 & S2 0109+22 & 14.5 & 1.078 & 5.818 & \nodata & 1, 2 \\
0109+224 & S2 0109+22 & 22 & 0.857 & 7.734 & \nodata & \nodata \\
0109+224 & S2 0109+22 & 37 & 1.918 & 2.943 & \nodata & \nodata \\
\tableline
0235+164 & AO 0235+164 & 4.8 & 0.961 & 1.848 & 11.523 & \nodata \\
0235+164 & AO 0235+164 & 8 & 0.606 & 1.848 & 5.615 & 2 \\
0235+164 & AO 0235+164 & 14.5 & 0.857 & 5.544 & 5.658 & \nodata \\
0235+164 & AO 0235+164 & 22 & 0.961 & 5.681 & 5.609 & \nodata \\
0235+164 & AO 0235+164 & 37 & 2.709 & 3.901 & \nodata & \nodata \\
0235+164 & AO 0235+164 & 90 & 0.763 & 1.848 & \nodata & \nodata \\
\tableline
0422+004 & PKS 0422+004 & 4.8 & $\geq 10.784$ & 7.050 & \nodata & 1, 2 \\
0422+004 & PKS 0422+004 & 8 & $\geq 10.784$ & 7.871 & \nodata & 1 \\
0422+004 & PKS 0422+004 & 14.5 & 3.826 & 6.092 & \nodata & 1 \\
0422+004 & PKS 0422+004 & 22 & $\geq 9.612$ & \nodata & \nodata & \nodata \\
0422+004 & PKS 0422+004 & 37 & 2.709 & \nodata & \nodata & \nodata \\
\tableline
0716+714 & S5 0716+714 & 4.8 & 0.961 & 5.818 & 5.584 & \nodata \\
0716+714 & S5 0716+714 & 8 & $\geq 8.566$ & 5.270 & 4.213 & \nodata \\
0716+714 & S5 0716+714 & 14.5 & 0.192 & 1.985 & \nodata & \nodata \\
0716+714 & S5 0716+714 & 22 & $\geq 12.100$ & 1.711 & \nodata & 2 \\
0716+714 & S5 0716+714 & 37 & 0.680 & 2.259 & 2.232 & 2 \\
0716+714 & S5 0716+714 & 90 & $\geq 6.805$ & \nodata & \nodata & \nodata \\
\tableline
0735+178 & PKS 0735+178 & 4.8 & 3.826 & \nodata & \nodata & \nodata \\
0735+178 & PKS 0735+178 & 8 & 5.405 & \nodata & 14.105 & \nodata \\
0735+178 & PKS 0735+178 & 14.5 & 3.826 & \nodata & \nodata & \nodata \\
0735+178 & PKS 0735+178 & 22 & 6.065 & \nodata & \nodata & \nodata \\
0735+178 & PKS 0735+178 & 37 & 2.152 & \nodata & \nodata & \nodata \\
0735+178 & PKS 0735+178 & 90 & 2.709 & \nodata & \nodata & \nodata \\
\tableline
0754+100 & PKS 0754+100 & 4.8 & 2.414 & 7.050 & 10.037 & \nodata \\
0754+100 & PKS 0754+100 & 8 & 3.039 & 7.871 & 10.814 & 2 \\
0754+100 & PKS 0754+100 & 14.5 & 4.293 & 10.883 & 10.628 & \nodata \\
0754+100 & PKS 0754+100 & 22 & 0.541 & 3.354 & 6.255 & \nodata \\
0754+100 & PKS 0754+100 & 37 & 1.210 & 2.806 & 2.819 & 2 \\
\tableline
0814+425 & OJ 425 & 4.8 & $\geq 10.784$ & \nodata & \nodata & \nodata \\
0814+425 & OJ 425 & 8 & 0.606 & \nodata & 14.070 & \nodata \\
0814+425 & OJ 425 & 14.5 & 7.635 & \nodata & \nodata & \nodata \\
0814+425 & OJ 425 & 22 & $\geq 5.405$ & \nodata & \nodata & \nodata \\
0814+425 & OJ 425 & 37 & $\geq 4.293$ & \nodata & \nodata & \nodata \\
\tableline
0851+202 & OJ 287 & 4.8 & 0.341 & \nodata & \nodata & \nodata \\
0851+202 & OJ 287 & 8 & 0.215 & 6.502 & \nodata & 2 \\
0851+202 & OJ 287 & 14.5 & 0.241 & \nodata & \nodata & \nodata \\
0851+202 & OJ 287 & 22 & 0.304 & \nodata & \nodata & \nodata \\
0851+202 & OJ 287 & 37 & 0.482 & 5.818 & \nodata & \nodata \\
0851+202 & OJ 287 & 90 & 0.241 & 5.133 & 1.031 & 2 \\
0851+202 & OJ 287 & 230 & \nodata & 4.038 & \nodata & 2 \\
\tableline
0954+658 & S4 0954+65 & 4.8 & 10.784 & \nodata & \nodata & \nodata \\
0954+658 & S4 0954+65 & 8 & 3.039 & 5.955 & \nodata & 1 \\
0954+658 & S4 0954+65 & 14.5 & 12.100 & 3.217 & \nodata & 1 \\
0954+658 & S4 0954+65 & 22 & 1.210 & 2.806 & \nodata & 1 \\
0954+658 & S4 0954+65 & 37 & 0.383 & 2.533 & \nodata & 1 \\
\tableline
1101+384 & Mark 421 & 4.8 & 0.215 & 1.164 & \nodata & 1 \\
1101+384 & Mark 421 & 8 & 0.271 & 1.848 & 5.236 & 1 \\
1101+384 & Mark 421 & 14.5 & 0.241 & 2.806 & 9.604 & 1, 2 \\
1101+384 & Mark 421 & 22 & 0.271 & 0.890 & 0.800 & 1, 2 \\
1101+384 & Mark 421 & 37 & \nodata & 1.848 & \nodata & 1, 2 \\
\tableline
1147+245 & B2 1147+24 & 8 & \nodata & 4.860 & \nodata & 1 \\
1147+245 & B2 1147+24 & 22 & \nodata & 3.354 & \nodata & 1 \\
1147+245 & B2 1147+24 & 37 & \nodata & \nodata & \nodata & \nodata \\
\tableline
1219+285 & ON 231 & 4.8 & 6.805 & 10.609 & \nodata & \nodata \\
1219+285 & ON 231 & 8 & $\geq 24.143$ & \nodata & \nodata & \nodata \\
1219+285 & ON 231 & 14.5 & $\geq 24.143$ & \nodata & \nodata & \nodata \\
1219+285 & ON 231 & 22 & $\geq 21.518$ & 7.324 & 7.962 & \nodata \\
1219+285 & ON 231 & 37 & 4.293 & 7.324 & 8.768 & \nodata \\
\tableline
1308+326 & AUCVn & 4.8 & 3.410 & 11.704 & \nodata & \nodata \\
1308+326 & AUCVn & 8 & 4.293 & 10.746 & 14.594 & \nodata \\
1308+326 & AUCVn & 14.5 & 3.039 & 10.335 & 8.890 & \nodata \\
1308+326 & AUCVn & 22 & 2.709 & \nodata & \nodata & \nodata \\
1308+326 & AUCVn & 37 & 3.039 & 10.472 & 12.002 & \nodata \\
1308+326 & AUCVn & 90 & \nodata & 3.765 & \nodata & 3 \\
\tableline
1413+135 & PKS 1413+135 & 4.8 & 2.152 & 9.377 & 7.618 & 1 \\
1413+135 & PKS 1413+135 & 8 & 2.709 & 9.377 & 8.206 & \nodata \\
1413+135 & PKS 1413+135 & 14.5 & 2.709 & 6.776 & 9.109 & \nodata \\
1413+135 & PKS 1413+135 & 22 & 1.709 & 3.901 & 7.682 & \nodata \\
1413+135 & PKS 1413+135 & 37 & 1.523 & 3.901 & 8.067 & \nodata \\
1413+135 & PKS 1413+135 & 90 & \nodata & 2.259 & \nodata & \nodata \\
\tableline
1418+546 & OQ 530 & 4.8 & 1.210 & 3.901 & \nodata & \nodata \\
1418+546 & OQ 530 & 8 & 1.078 & 2.533 & \nodata & 2 \\
1418+546 & OQ 530 & 14.5 & 1.523 & 4.723 & \nodata & 2 \\
1418+546 & OQ 530 & 22 & $\geq 7.635$ & 4.175 & \nodata & \nodata \\
1418+546 & OQ 530 & 37 & 0.606 & 7.050 & \nodata & \nodata \\
\tableline
1538+149 & 4C 14.60 & 4.8 & 5.405 & 5.133 & \nodata & 1 \\
1538+149 & 4C 14.60 & 8 & 2.152 & \nodata & \nodata & \nodata \\
1538+149 & 4C 14.60 & 14.5 & 1.358 & \nodata & \nodata & \nodata \\
1538+149 & 4C 14.60 & 22 & 7.635 & \nodata & \nodata & \nodata \\
1538+149 & 4C 14.60 & 37 & 7.635 & 6.913 & \nodata & 1 \\
\tableline
1652+398 & Mark 501 & 4.8 & 2.709 & 4.038 & \nodata & 1 \\
1652+398 & Mark 501 & 8 & 4.817 & \nodata & \nodata & \nodata \\
1652+398 & Mark 501 & 14.5 & 3.039 & 10.746 & \nodata & 1 \\
1652+398 & Mark 501 & 22 & $\geq 12.100$ & 1.300 & 4.958 & 1 \\
1652+398 & Mark 501 & 37 & $\geq 12.100$ & 3.354 & 8.454 & 1 \\
\tableline
1749+096 & PKS 1749+096 & 4.8 & 1.358 & 4.449 & \nodata & \nodata \\
1749+096 & PKS 1749+096 & 8 & 2.152 & 6.776 & 6.956 & \nodata \\
1749+096 & PKS 1749+096 & 14.5 & 1.078 & 2.533 & 3.043 & \nodata \\
1749+096 & PKS 1749+096 & 22 & 0.606 & 2.806 & 3.055 & \nodata \\
1749+096 & PKS 1749+096 & 37 & 0.341 & 2.122 & 9.813 & 2 \\
1749+096 & PKS 1749+096 & 90 & 3.040 & 1.300 & 1.944 & 2 \\
\tableline
1803+784 & S5 1803+784 & 4.8 & 0.857 & 10.883 & 10.849 & \nodata \\
1803+784 & S5 1803+784 & 8 & 1.918 & 2.943 & \nodata & 2 \\
1803+784 & S5 1803+784 & 14.5 & 2.152 & 4.312 & 9.785 & \nodata \\
1803+784 & S5 1803+784 & 22 & \nodata & 4.723 & \nodata & \nodata \\
1803+784 & S5 1803+784 & 37 & \nodata & 2.806 & \nodata & \nodata \\
\tableline
1807+698 & 3C 371.0 & 4.8 & 3.826 & \nodata & \nodata & \nodata \\
1807+698 & 3C 371.0 & 8 & 10.784 & 9.240 & \nodata & 2 \\
1807+698 & 3C 371.0 & 14.5 & 4.817 & \nodata & \nodata & \nodata \\
1807+698 & 3C 371.0 & 22 & \nodata & 1.300 & \nodata & \nodata \\
1807+698 & 3C 371.0 & 37 & \nodata & 2.533 & \nodata & 1 \\
\tableline
1823+568 & 4C 56.27 & 4.8 & 3.039 & 9.377 & \nodata & \nodata \\
1823+568 & 4C 56.27 & 8 & 3.039 & 7.871 & \nodata & \nodata \\
1823+568 & 4C 56.27 & 14.5 & 1.918 & 2.806 & \nodata & \nodata \\
1823+568 & 4C 56.27 & 22 & 4.293 & \nodata & \nodata & \nodata \\
1823+568 & 4C 56.27 & 37 & 2.709 & \nodata & \nodata & \nodata \\
\tableline
2007+776 & S5 2007+77 & 4.8 & 1.523 & 3.354 & \nodata & \nodata \\
2007+776 & S5 2007+77 & 8 & 0.680 & 2.259 & \nodata & \nodata \\
2007+776 & S5 2007+77 & 14.5 & 1.709 & \nodata & \nodata & \nodata \\
2007+776 & S5 2007+77 & 22 & 3.826 & \nodata & \nodata & \nodata \\
2007+776 & S5 2007+77 & 37 & \nodata & 3.080 & 3.061 & \nodata \\
\tableline
2200+420 & BL Lac & 4.8 & 3.826 & 7.461 & 8.514 & \nodata \\
2200+420 & BL Lac & 8 & 0.763 & 2.806 & 3.803 & \nodata \\
2200+420 & BL Lac & 14.5 & 0.961 & 7.461 & 8.267 & \nodata \\
2200+420 & BL Lac & 22 & 2.414 & 7.734 & 8.730 & 2 \\
2200+420 & BL Lac & 37 & 0.482 & 3.491 & 8.481 & 2 \\
2200+420 & BL Lac & 90 & \nodata & 4.449 & 5.984 & \nodata \\
\tableline
2223-052 & 3C 446 & 4.8 & 3.410 & 9.103 & 11.934 & \nodata \\
2223-052 & 3C 446 & 8 & 1.523 & 8.419 & 15.241 & \nodata \\
2223-052 & 3C 446 & 14.5 & 2.709 & 11.157 & 5.710 & \nodata\\
2223-052 & 3C 446 & 22 & 2.709 & \nodata & \nodata & \nodata \\
2223-052 & 3C 446 & 37 & 3.039 & \nodata & \nodata & \nodata\\
2223-052 & 3C 446 & 90 & 1.358 & 5.955 & \nodata & \nodata\\
\enddata
\tablecomments{1 = faint source, 2 = multiple time scales, 3 = exceptionally large errors in $T_{DCF}$. See text for details.}
\end{deluxetable}


\begin{deluxetable}{l l c c c c c c}
\tablecaption{The number of flares included in the analysis and the mean values of the duration, rise time, decay time, absolute peak flux and relative peak flux for the 13 flaring sources in our sample. The parameters have been calculated for all frequency bands and separately for 37 GHz.\label{source_means}}
\tablewidth{0pt}
\tabletypesize{\small}
\tablehead{\colhead{} & \colhead{} & \colhead{Number} & \colhead{Duration} & \colhead{Rise} & \colhead{Decay} & \colhead{Absolute} & \colhead{Relative}\\
\colhead{Source} & \colhead{Frequency} & \colhead{of flares} & \colhead{[yr]} & \colhead{time [yr]} & \colhead{time [yr]} & \colhead{peak flux [Jy]} & \colhead{peak flux [Jy]}}
\tablecolumns{8}
\startdata
S2 0109+22 & All & 3 & 3.9 & 1.3 & 2.6 & 1.6 & 1.1 \\
 & 37 GHz & 3 & 3.7 & 1.0 & 2.7 & 2.3 & 1.7 \\
AO 0235+164 & All & 4 & 2.3 & 1.2 & 1.1 & 4.5 & 3.6 \\
 & 37 GHz & 4 & 2.4 & 1.4 & 1.1 & 5.3 & 4.2 \\
PKS 0422+004 & All & 2 & 3.3 & 1.5 & 1.7 & 1.5 & 1.1 \\
 & 37 GHz & 2 & 2.0 & 0.6 & 1.3 & 1.9 & 1.3 \\
S5 0716+714 & All & 2 & 3.0 & 1.2 & 1.8 & 2.9 & 2.4 \\
 & 37 GHz & 2 & 3.1 & 0.9 & 2.2 & 4.4 & 4.1 \\
PKS 0735+178 & All & 1 & 10.8 & 2.9 & 7.8 & 4.6 & 3.5 \\
 & 37 GHz & 1 & 10.7 & 1.9 & 8.8 & 5.3 & 4.4 \\
PKS 0754+100 & All & 2 & 3.4 & 1.5 & 2.0 & 2.3 & 1.3 \\
 & 37 GHz & 2 & 3.3 & 1.0 & 2.2 & 2.7 & 1.7 \\
OJ 287 & All & 9 & 1.4 & 0.7 & 0.7 & 4.9 & 2.7 \\
 & 37 GHz & 9 & 1.3 & 0.7 & 0.6 & 5.9 & 3.5 \\
1308+326 & All & 2 & 12.8 & 5.9 & 6.9 & 4.1 & 3.5 \\
 & 37 GHz & 2 & 13.2 & 3.9 & 8.1 & 3.5 & 2.9 \\
PKS 1413+135 & All & 2 & 4.8 & 2.2 & 2.5 & 3.0 & 2.2 \\
 & 37 GHz & 2 & 4.8 & 2.6 & 2.2 & 3.5 & 3.0 \\
PKS 1749+096 & All & 5 & 1.7 & 0.9 & 0.8 & 6.4 & 4.3 \\
 & 37 GHz & 5 & 1.7 & 0.8 & 0.9 & 7.5 & 5.2 \\
S5 2007+77 & All & 1 & 3.3 & 1.5 & 1.8 & 3.3 & 2.1 \\
 & 37 GHz & 1 & 3.3 & 1.5 & 1.9 & 3.0 & 2.2 \\
BL Lac & All & 9 & 1.8 & 0.9 & 0.9 & 5.0 & 2.9 \\
 & 37 GHz & 9 & 1.6 & 0.8 & 0.8 & 5.1 & 3.2 \\
3C 446 & All & 3 & 5.8 & 2.7 & 3.1 & 7.8 & 4.6 \\
 & 37 GHz & 3 & 5.4 & 3.1 & 2.3 & 8.1 & 5.4 \\
\enddata
\end{deluxetable}


\begin{table}
\caption{Minimum, maximum, mean and median values of flare duration and absolute and relative peak fluxes (both $S_\mathrm{max}-S_\mathrm{min}$ and $S_\mathrm{max}/S_\mathrm{min}$) for all frequencies used in the analysis. Values are also shown for duration and absolute flux of the flares normalized to the values at 22\,GHz.\label{parameters}}
\begin{tabular}{l c c c c | c c c c}
$\nu$ [GHz] & \multicolumn{4}{c} {Duration [yr]} & \multicolumn{4}{c} {Normalized duration}\\
 & min. & max. & mean & median  & min. & max. & mean & median\\
\tableline
4.8 & 0.6 & 12.7 & 3.5 & 2.7 & 0.4 & 2.2 & 1.1 & 1.0\\
8 & 0.7 & 12.8 & 3.4 & 2.7 & 0.4 & 3.2 & 1.1 & 1.1\\
14.5 & 0.3 & 12.4 & 3.0 & 2.4 & 0.7 & 3.0 & 1.1 & 1.0\\
22 & 0.3 & 13.0 & 2.9 & 2.3 & 1.0 & 1.0 & 1.0 & 1.0\\
37 & 0.4 & 13.2 & 2.9 & 2.4 & 0.7 & 1.9 & 1.0 & 1.0\\
90 & 0.6 & 10.3 & 3.1 & 2.3 & 0.7 & 1.7 & 1.0 & 1.0\\
230 & 0.9 & 9.9 & 3.4 & 2.3 & 0.6 & 1.1 & 0.9 & 0.9\\
\tableline
$\nu$ [GHz] & \multicolumn{4}{c} {Absolute peak flux [Jy]} & \multicolumn{4}{c} {Normalized abs. peak flux}\\
& min. & max. & mean & median & min. & max. & mean & median\\
\tableline
4.8 & 0.7 & 7.1 & 3.4 & 3.2 & 0.28  & 1.08 & 0.67 & 0.63\\
8 & 0.9 & 8.0 & 4.0 & 4.3 & 0.45 & 1.08 & 0.80 & 0.81\\
14.5 & 0.8 & 9.0 & 4.5 & 4.2 & 0.53 & 1.12 & 0.90 & 0.91\\
22 & 1.2 & 10.7 & 5.0 & 4.9 & 1.0 & 1.0 & 1.0 & 1.0\\
37 & 1.4 & 10.9 & 5.1 & 5.1 & 0.81 & 1.28 & 1.03 & 1.04\\
90 & 1.3 & 12.1 & 5.1 & 4.4 & 0.56 & 1.89 & 0.98 & 0.90\\
230 & 1.2 & 11.7 & 4.4 & 3.2 & 0.42 &  1.09 & 0.70 & 0.67\\
\tableline
$\nu$ [GHz] & \multicolumn{4}{c} {Relative peak flux [Jy]} & \multicolumn{4}{c} {Relative peak flux ($S_\mathrm{max}/S_\mathrm{min}$)}\\
& min. & max. & mean & median  & min. & max. & mean & median\\
\tableline
4.8 & 0.4 & 4.8 & 1.9 & 1.7 & 1.4 & 17.8 & 3.1 & 2.2\\
8 & 0.7 & 5.1 & 2.4 & 2.1 & 1.4 & 16.7 & 3.7 & 2.6\\
14.5 & 0.8 & 5.8 & 2.8 & 2.4 & 1.5 & 18.4 & 3.9 & 2.6\\
22 & 0.8 & 7.1 & 3.2 & 2.8 & 1.8 & 11.1 & 3.6 & 2.8\\
37 & 1.0 & 7.6 & 3.5 & 3.1 & 1.8 & 18.5 & 4.4 & 3.3\\
90 & 0.6 & 9.4 & 3.5 & 2.8 & 1.6 & 12.2 & 4.0 & 2.9\\
230 & 0.9 & 10.1 & 3.3 & 1.9 & 1.8 & 7.1 & 4.2 & 3.7\\
\tableline
\end{tabular}
\end{table}



\begin{figure}
\epsscale{1.0}
\figurenum{2.2}
\plotone{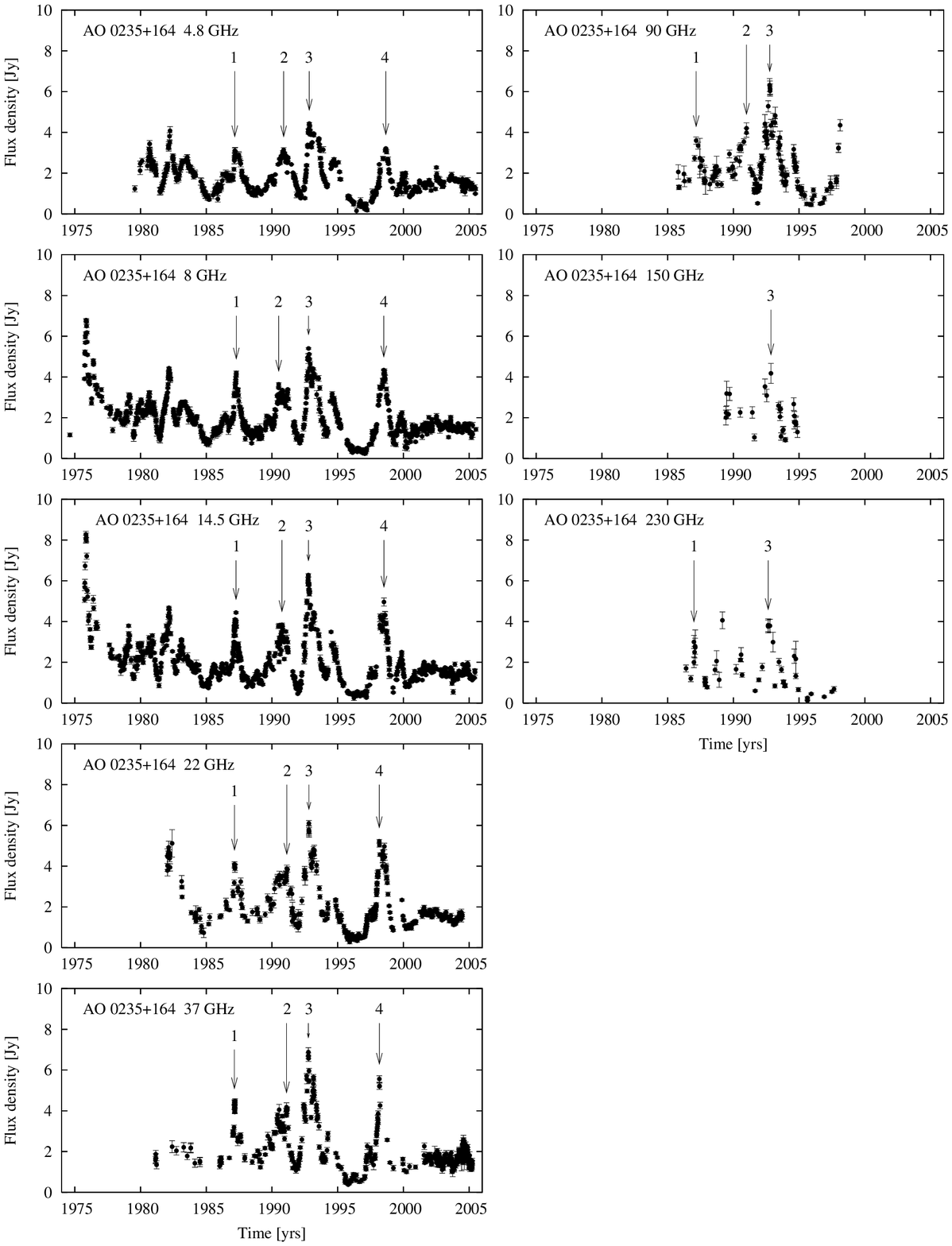}
\caption{Flux curves of objects in the flare analysis. The peak of each flare included in the analysis is marked in the curve.}
\end{figure}

\begin{figure}
\figurenum{2.3}
\plotone{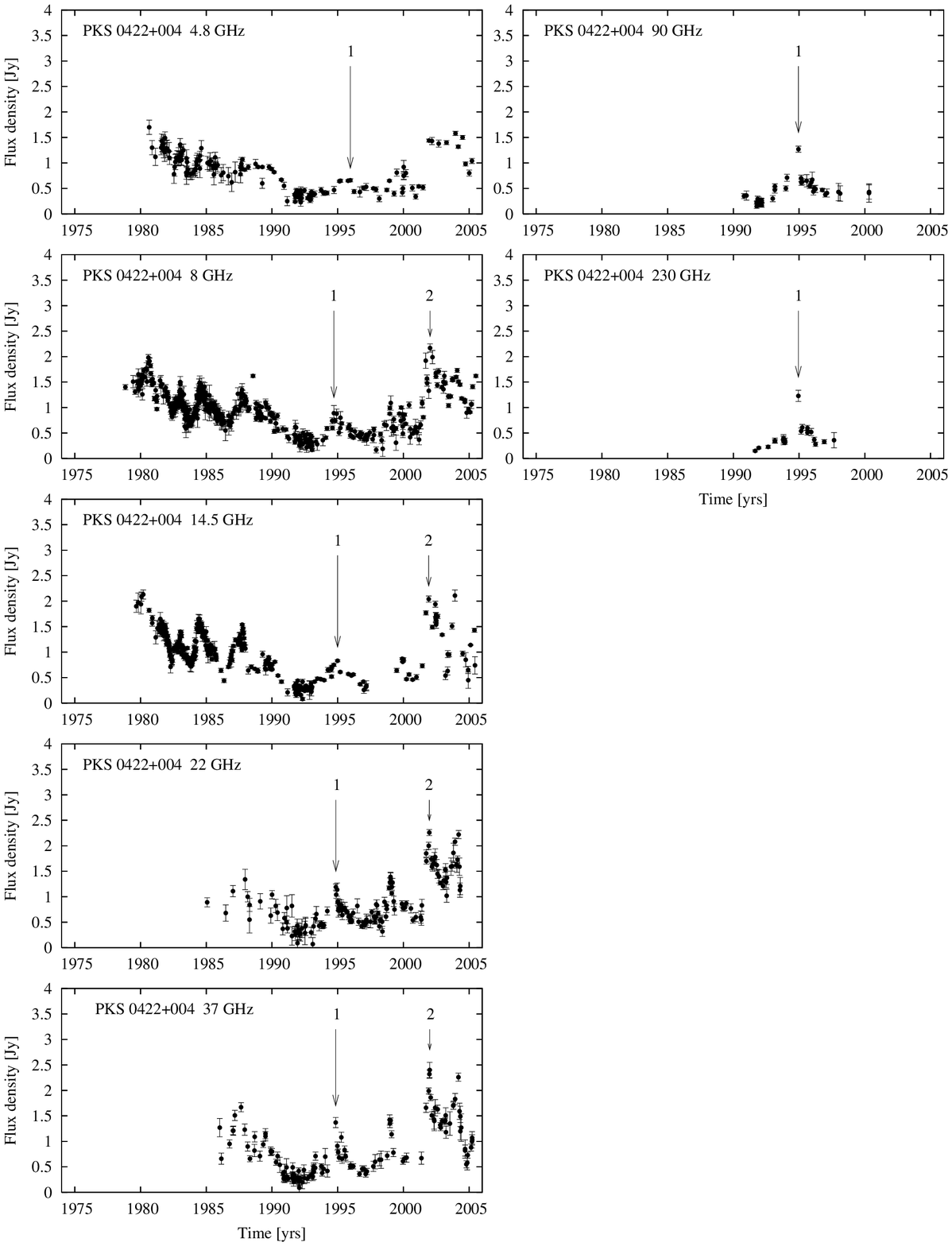}
\caption{Flux curves of objects in the flare analysis. The peak of each flare included in the analysis is marked in the curve.}
\end{figure}

\begin{figure}
\figurenum{2.4}
\epsscale{0.5}
\plotone{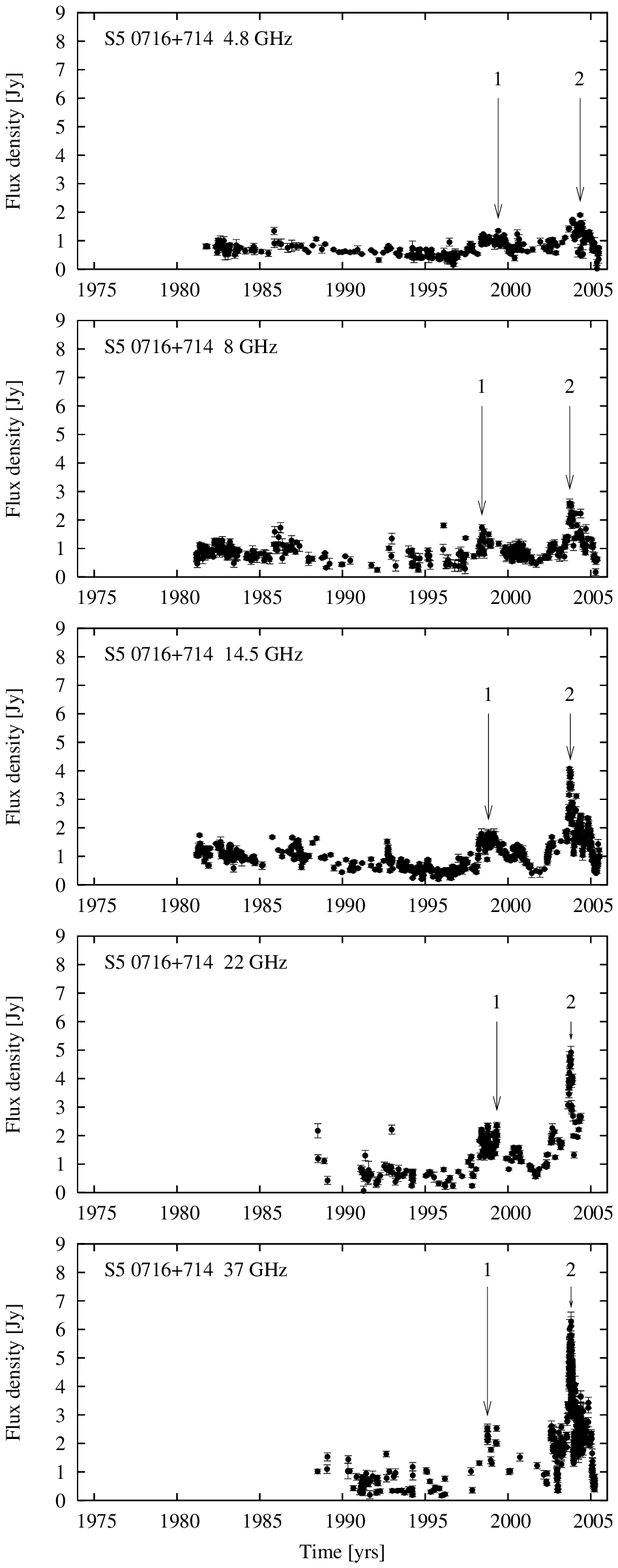}
\caption{Flux curves of objects in the flare analysis. The peak of each flare included in the analysis is marked in the curve.}
\end{figure}

\begin{figure}
\figurenum{2.5}
\epsscale{1.0}
\plotone{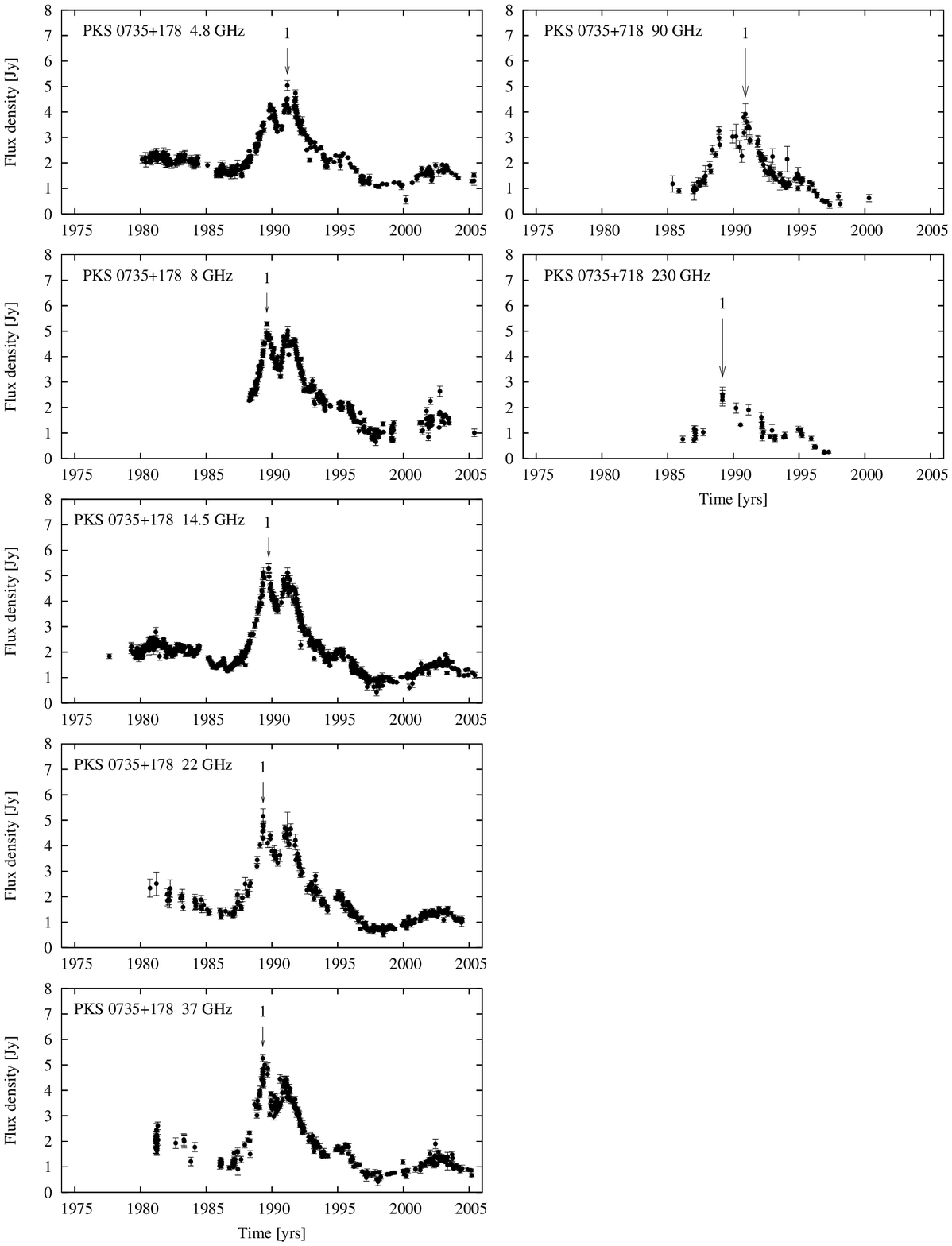}
\caption{Flux curves of objects in the flare analysis. The peak of each flare included in the analysis is marked in the curve.}
\end{figure}

\begin{figure}
\figurenum{2.6}
\plotone{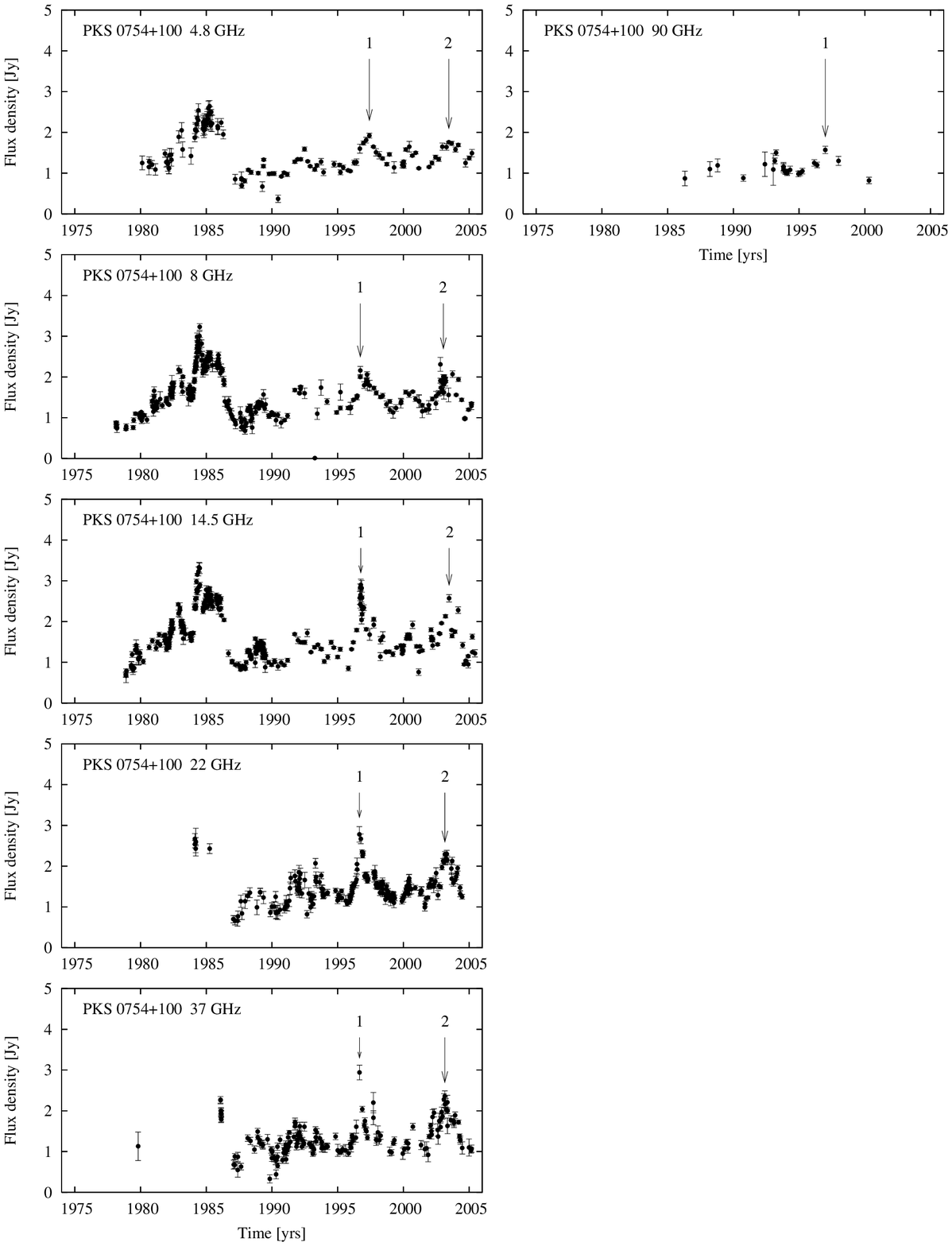}
\caption{Flux curves of objects in the flare analysis. The peak of each flare included in the analysis is marked in the curve.}
\end{figure}

\begin{figure}
\figurenum{2.7}
\plotone{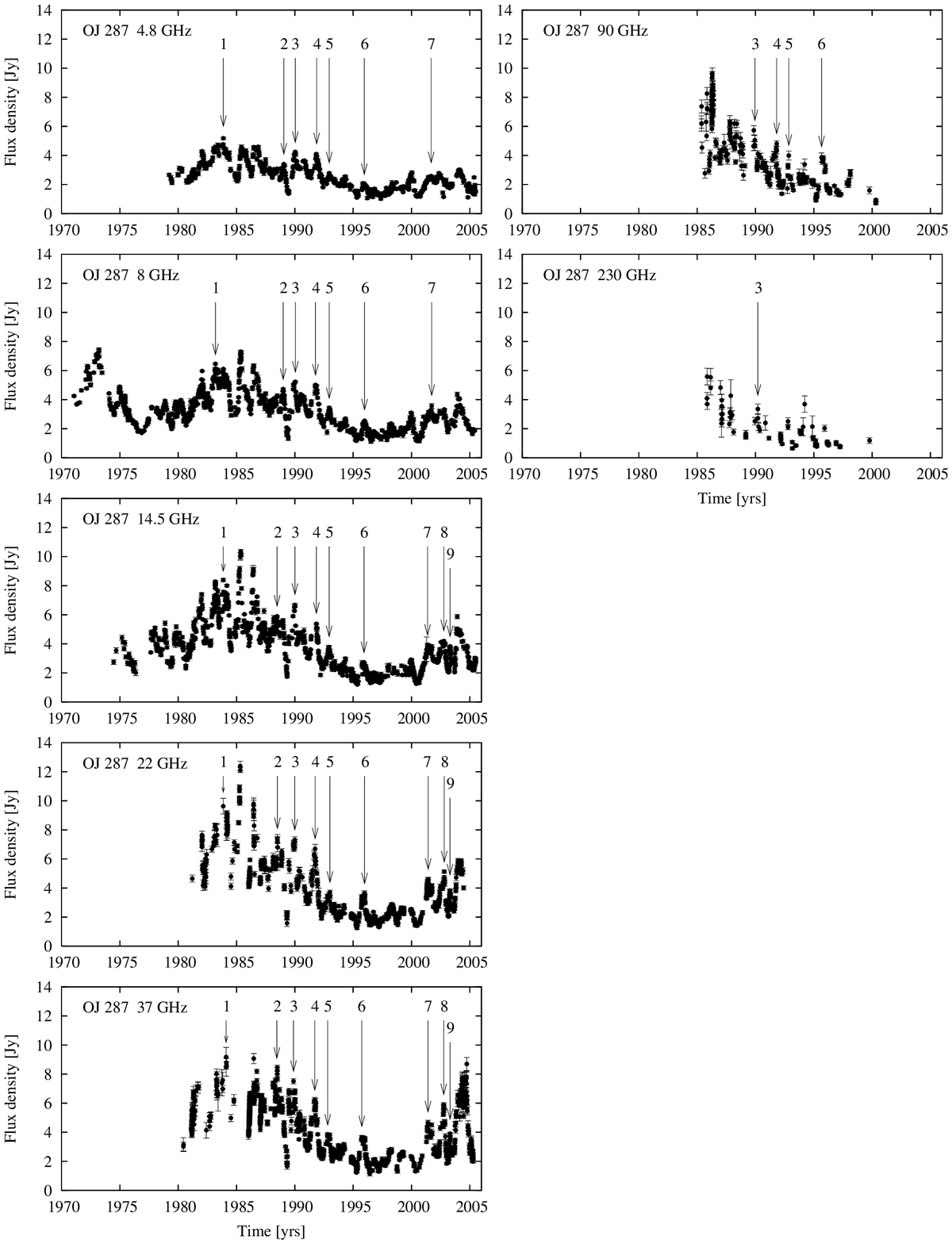}
\caption{Flux curves of objects in the flare analysis. The peak of each flare included in the analysis is marked in the curve.}
\end{figure}

\begin{figure}
\figurenum{2.8}
\epsscale{0.5}
\plotone{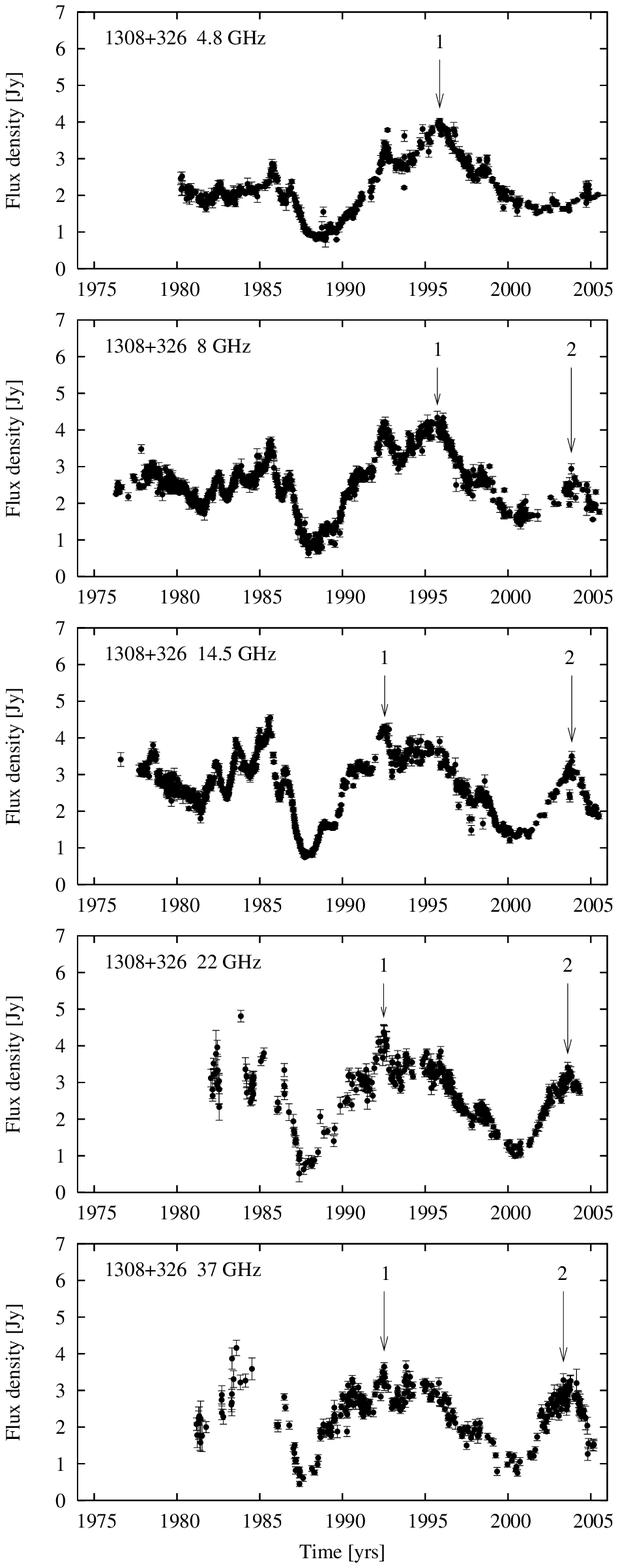}
\caption{Flux curves of objects in the flare analysis. The peak of each flare included in the analysis is marked in the curve.}
\end{figure}

\begin{figure}
\figurenum{2.9}
\epsscale{1.0}
\plotone{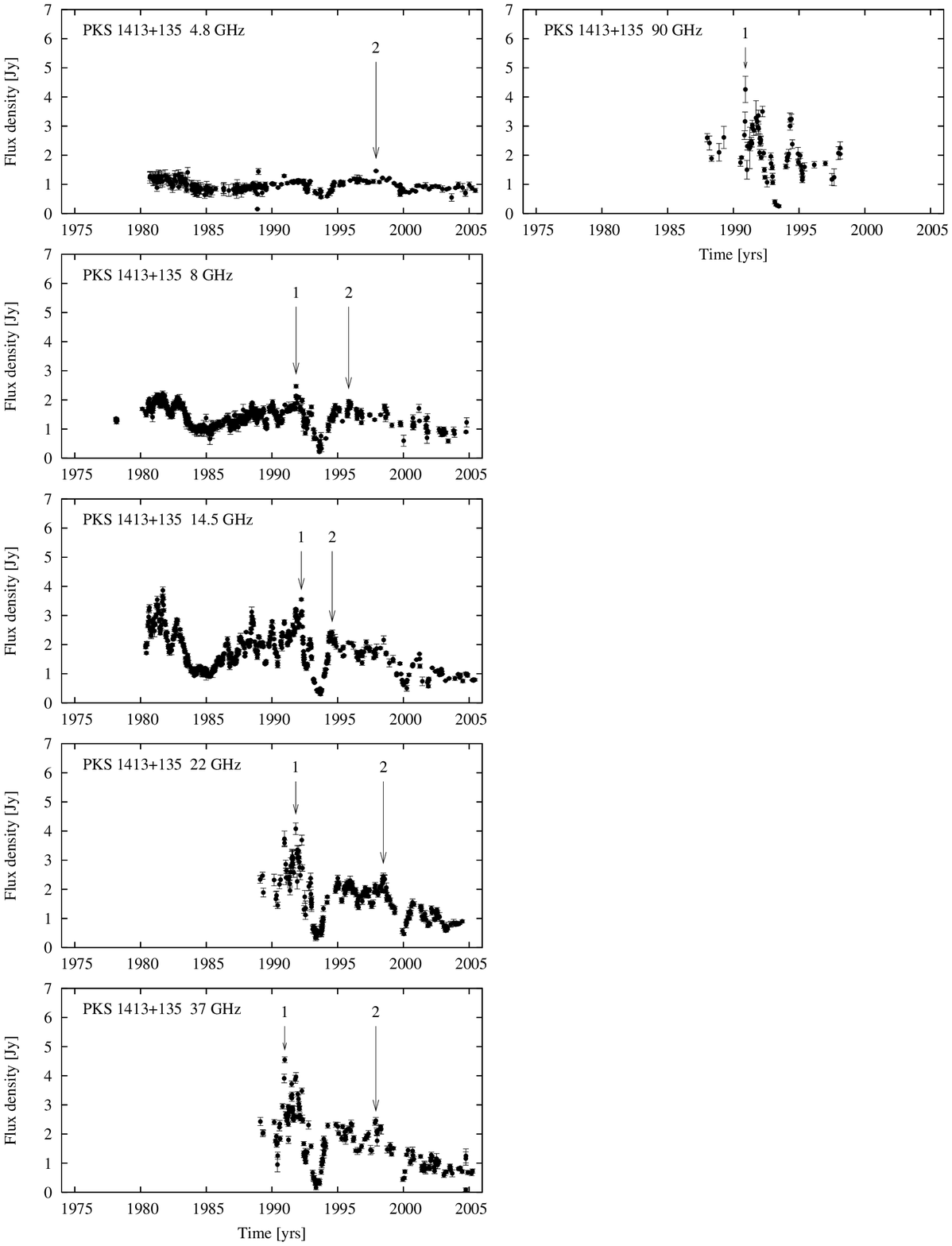}
\caption{Flux curves of objects in the flare analysis. The peak of each flare included in the analysis is marked in the curve.}
\end{figure}

\begin{figure}
\figurenum{2.10}
\plotone{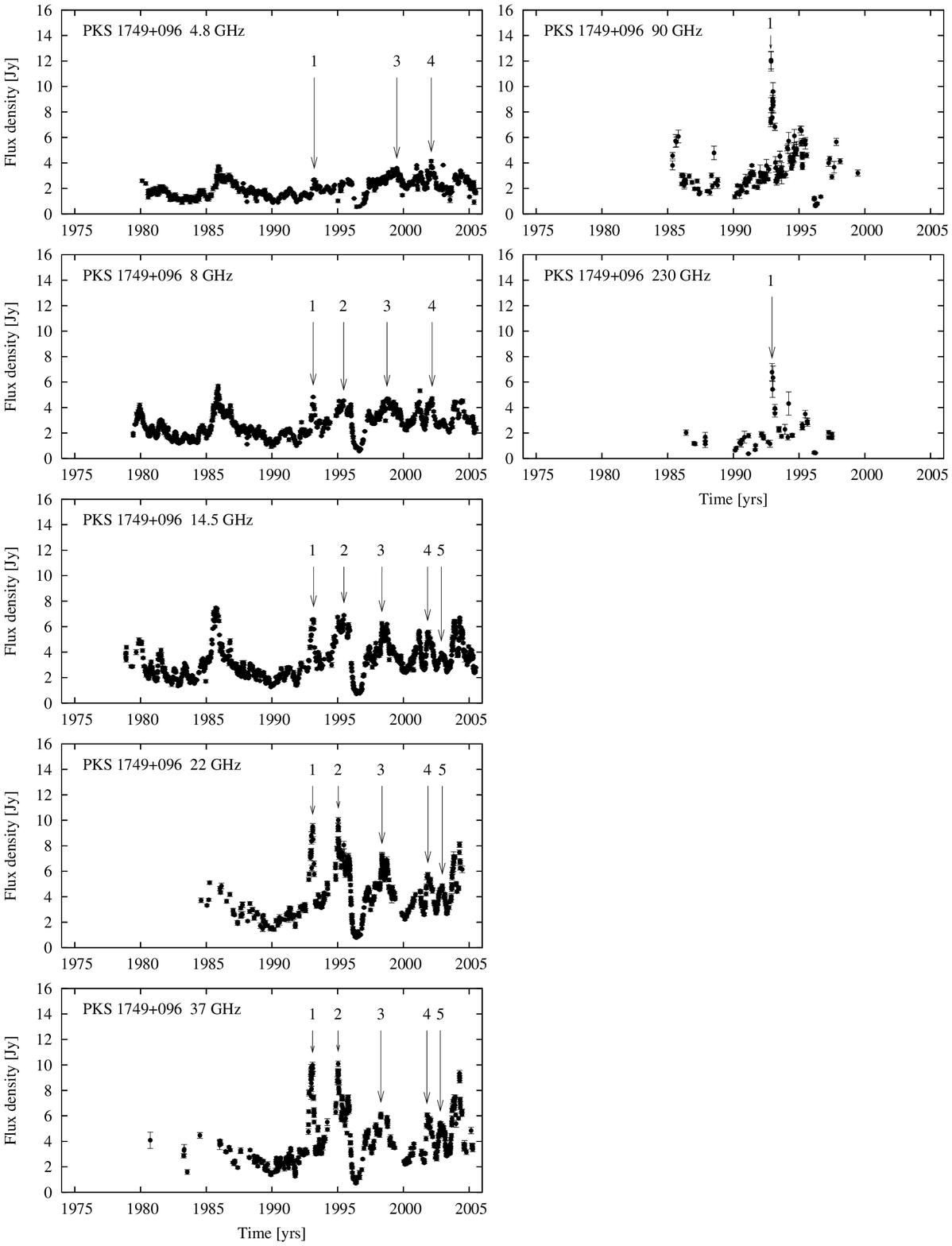}
\caption{Flux curves of objects in the flare analysis. The peak of each flare included in the analysis is marked in the curve.}
\end{figure}

\begin{figure}
\figurenum{2.11}
\epsscale{0.5}
\plotone{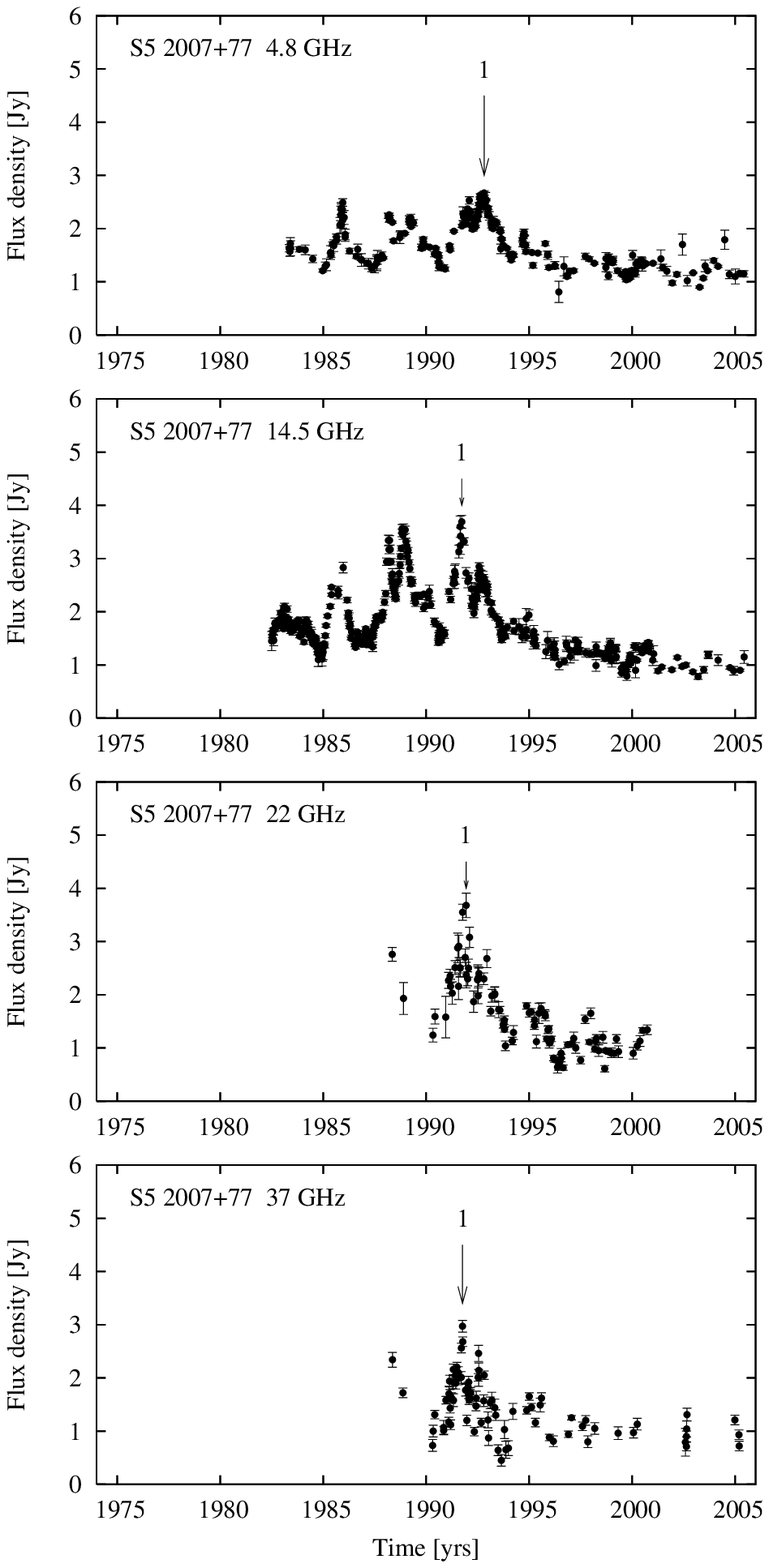}
\caption{Flux curves of objects in the flare analysis. The peak of each flare included in the analysis is marked in the curve.}
\end{figure}

\begin{figure}
\epsscale{1.0}
\figurenum{2.12}
\plotone{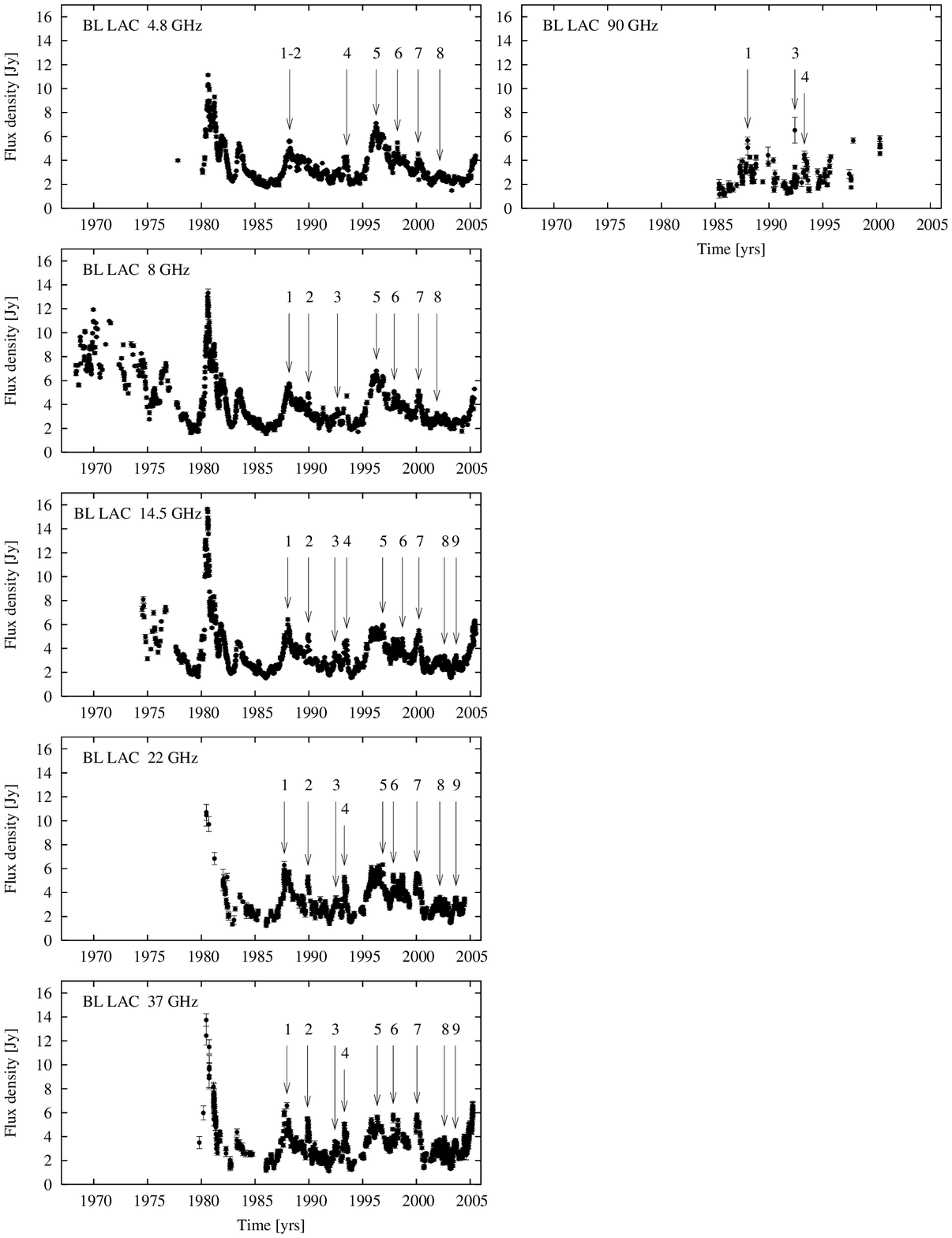}
\caption{Flux curves of objects in the flare analysis. The peak of each flare included in the analysis is marked in the curve.}
\end{figure}

\begin{figure}
\figurenum{2.13}
\plotone{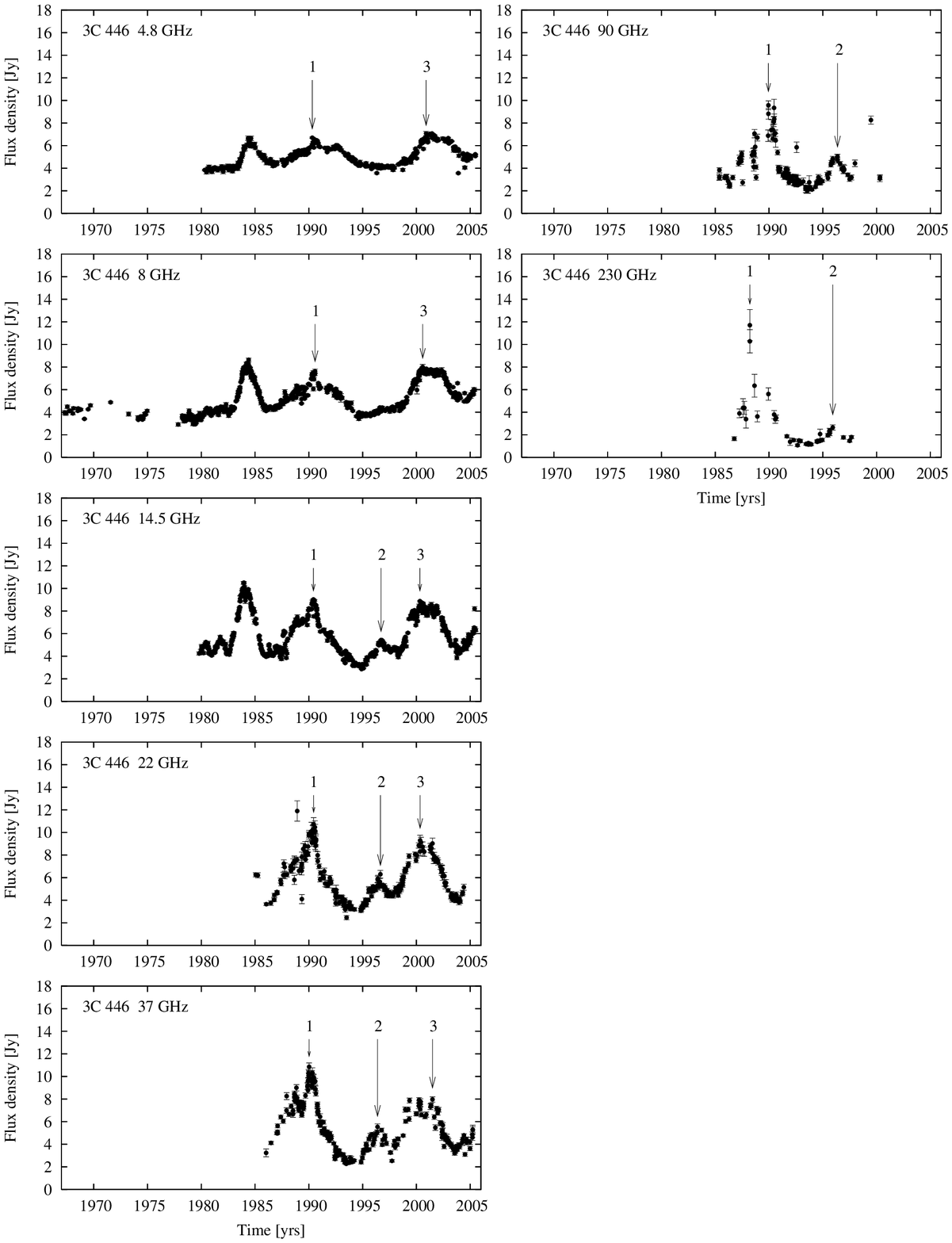}
\caption{Flux curves of objects in the flare analysis. The peak of each flare included in the analysis is marked in the curve.}
\end{figure}


\end{document}